\documentclass[twocolumn,3p]{elsarticle}

\usepackage{lineno,hyperref}
\hypersetup{colorlinks=true,citecolor=blue,linkcolor=blue,urlcolor=blue} % HYPERLINKS
\modulolinenumbers[5]

\usepackage{subfigure}
\usepackage{graphicx}  % needed for figures
\usepackage{dcolumn}   % needed for some tables
\usepackage{bm}		   % for math
\usepackage{amssymb}   % for math
\usepackage{soul}
\usepackage[dvipsnames]{xcolor}
\usepackage{colortbl}
\usepackage{longtable}

\newcommand{\mo}[1]{\textcolor{black}{#1}} % correction
\newcommand{\cn}[1]{\textcolor{black}{#1}}  % comment number 
\newcommand{\gh}[1]{\textcolor{black}{#1}} % correction
\newcommand{\co}[1]{\textcolor{black}{#1}} % correction

\setstcolor{red}

\journal{Acta Materialia}

\begin{document}

\begin{frontmatter}

\title{A computational high-throughput search for new ternary superalloys}

\author[mymainaddress]{Chandramouli Nyshadham}
\author[mysecondaryaddress]{Corey Oses}
\address[mymainaddress]{Department of Physics and Astronomy, Brigham Young University, Provo, Utah 84602, USA.}
\address[mysecondaryaddress]{Center for Materials Genomics, Duke University, Durham, North Carolina 27708, USA.}
\author[mymainaddress]{Jacob E. Hansen}


\author[myfourthaddress]{Ichiro Takeuchi}
\address[myfourthaddress]{Department of Materials Science and Engineering, University of Maryland, College Park, Maryland 20742, USA.}
\author[mysecondaryaddress,mythirdaddress]{Stefano Curtarolo} %\email{stefano@duke.edu}
\address[mythirdaddress]{Department of Mechanical Engineering and Materials Science and Department of Physics, Duke University, Durham, North Carolina 27708, USA.}
\author[mymainaddress]{Gus L. W. Hart\corref{mycorrespondingauthor}}
\cortext[mycorrespondingauthor]{Corresponding author.~Tel.:~+1-801-422-7444}
\ead{gus.hart@gmail.com}

 \begin{abstract}
In 2006, a novel cobalt-based superalloy was discovered~\cite{Sato_Science_2006} with
mechanical properties better than some conventional nickel-based superalloys. 
As with conventional superalloys, its high performance arises from the precipitate-hardening effect of a
coherent L1$_2$ phase, which is in two-phase equilibrium with the fcc matrix.
Inspired by this unexpected discovery of an L1$_2$ ternary phase, we performed a first-principles search
through 2224 ternary metallic systems for analogous precipitate-hardening phases of the
form $X_{3}$[$A_{0.5}, B_{0.5}$], where $X$ = Ni, Co, or Fe, and [$A,B$] = Li, Be, Mg, Al, Si,
Ca, Sc, Ti, V, Cr, Mn, Fe, Co, Ni, Cu, Zn Ga, Sr, Y, Zr, Nb, Mo, Tc, Ru, Rh, Pd, Ag, Cd, In, Sn,
Sb, Hf, Ta, W, Re, Os, Ir, Pt, Au, Hg, or Tl.  
We found 102 systems that have a smaller decomposition energy and a
lower formation enthalpy than the Co$_{3}$(Al, W) superalloy.
They have a stable two-phase equilibrium with the host matrix within the concentration range $0<x<1$
($X_{3}$[$A_{x}, B_{1-x}$]) and have a relative lattice mismatch with the host matrix of less
than or equal to 5\%.
These new candidates, narrowed from 2224 systems, suggest possible
experimental exploration for identifying new superalloys. 
Of these 102 systems, 37 are new; they have no reported phase diagrams in standard databases.
Based on cost, experimental difficulty, and toxicity, we limit these 37 to a shorter list of
six promising candidates of immediate interest.  
Our calculations are consistent with current
experimental literature where data exists.
   
\end{abstract}


\end{frontmatter}

%\linenumbers

\newpage

\section{Introduction}

Materials  scientists have developed large experimental databases of known materials over the last 
century~\cite{MatWeb,Matbase,ASMAlloyInternational,Pauling}.
Similar computational databases are being compiled by exploiting the power of supercomputers and advanced 
electronic structure methods~\cite{nmatHT,curtarolo:art65,curtarolo:art75,APL_Mater_Jain2013,Hachmann_JPCL_2011,nomad}.
The challenge now is to leverage the data to discover new materials by building computational
models~\cite{curtarolo:art49} and \cn{employing} 
machine learning \cn{methods}~\cite{Arsenault_ArX_2015,Hansen_JPCL_2015,curtarolo:art94,curtarolo:art85}.
Data mining and materials informatics approaches can also be used to identify structure/property relationships, 
which may suggest atomic combinations, stoichiometries, and structures not included in the database~\cite{curtarolo:art49}.

\begin{figure*}[tb]
\centering
\includegraphics[scale=0.75]{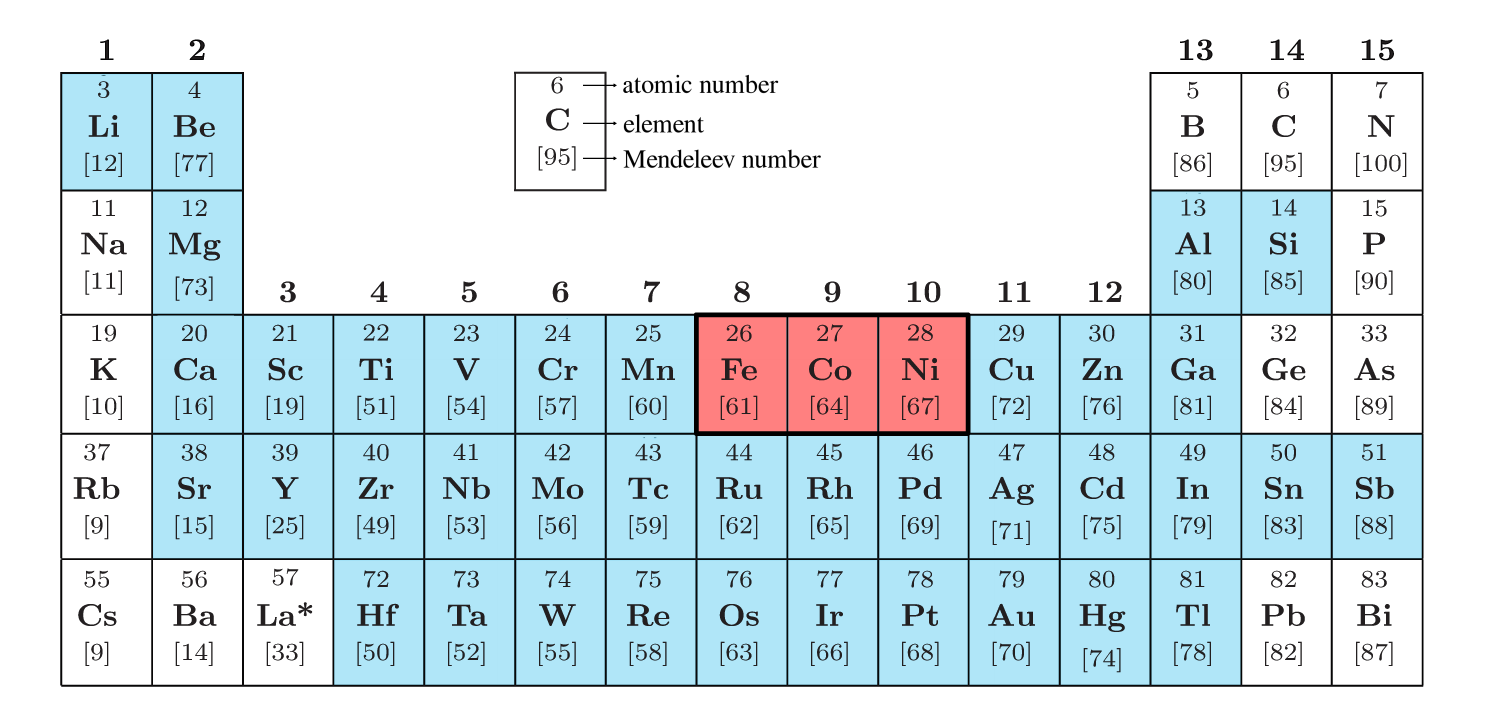}
\caption{For each base element in $X_{3}$[$A_{0.5}, B_{0.5}$], there are 40 elements chosen for $A$ and $B$, which includes 
38 elements (highlighted in blue) chosen from the periodic table and the remaining two of three base elements $X$ (highlighted in red).}
\label{ele}
\end{figure*} 
 
%But even simple searches through the data can identify new phases~\cite{curtarolo:art49}.
An emerging area in materials science is \cn{the} computational prediction of new materials using
high-throughput
approaches~\cite{nmatHT,curtarolo:art49,Arsenault_ArX_2015,Hansen_JPCL_2015,Ghiringhelli_PRL_2015,curtarolo:art63,curtarolo:art67,curtarolo:art57}.
Hundreds of thousands of hypothetical candidates can be explored much faster than by experimental
means.  In this work, a simple combinatorial search for ternary superalloys is performed in a
high-throughput fashion.  The extraordinary mechanical properties of superalloys at high
temperatures make them useful for many important applications in the aerospace and power generation
industries.  One of the basic traits of superalloys is that they generally occur in a
face-centered-cubic structure~\cite{Donachie_ASM_2002}.  The most common base elements for
superalloys are nickel, cobalt, and iron, but most are nickel-based.  In 2006, a new cobalt-based
superalloy, Co$_{3}$(Al, W), was confirmed to have better mechanical properties than many
nickel-based superalloys~\cite{Sato_Science_2006}.

This cobalt-based superalloy has the commonly occurring L1$_{2}$ phase
which creates coherent precipitates \cn{in the fcc matrix}.  A theoretical investigation of
Co$_{3}$(Al, W) was subsequently carried out by Saal and
Wolverton~\cite{Saal_ActMat_2013}.  \cn{To model the properties of the L1$_2$ solid solution
  phase observed at high temperature,} Saal and Wolverton used an L1$_{2}$-based
special quasirandom structure (SQS)~\cite{zunger_sqs}. In order to identify the
stoichiometry of the superalloy, they performed first-principles
calculations \cn{for solid solutions
Co$_{3}$[Al$_{x}$, W$_{1-x}$] with varying concentrations of Al and W.}
Their study includes finite temperature effects and point defect
energetics.  They showed that an L1$_{2}$-like random structure
with stoichiometry Co$_{3}$[Al$_{0.5}$, W$_{0.5}$] is consistent with
experiment.  Interestingly, their \cn{solid-solution-like} Co$_{3}$[Al$_{0.5}$,
  W$_{0.5}$] structure is metastable and predicted to have a
decomposition energy of 66 meV/atom (distance from the $T=0$ K convex
hull).  They show that high-temperature effects make this phase
thermodynamically competitive with other competing structures at
elevated temperatures.  The fact that a metastable structure
(Co$_{3}$[Al$_{0.5}$, W$_{0.5}$]) with a decomposition energy as high
as 66 meV/atom \mo{at T=0 K,} is competitive with many
commercially available superalloys \mo{at higher temperatures}
motivates our search for similar ternary systems containing an
L1$_{2}$-like solid solution phase.

Ideally, a computational search over potential superalloys would model
actual engineering observables (e.g., hardness) and \cn{consider} the
influence of small concentrations of impurities, finite temperature
effects, influence of vacancies, effects of polycrystallinity,
etc. Unfortunately, such calculations are extremely challenging even
for a single material and impractical for thousands of
candidate systems as in this work.

\gh{In known superalloy systems, L1$_2$-based
phases have large negative formation enthalpies, a small decomposition
energy, and a relatively small lattice mismatch between the host
matrix and the precipitate phase. Our search is for new ternary
systems with these same metrics. We further screen candidate alloy
systems for L1$_2$ precipitates either in  two-phase
equilibrium with the host matrix or likely to precipitate as
metastable phases.  Based on the relative lattice mismatch between the
host element and the precipitate phases any compound with a relative
lattice mismatch of $> 5\%$ is excluded.}

Using the \cn{solid-solution-like} structure identified by Saal and
Wolverton~\cite{Saal_ActMat_2013}, we performed an extensive combinatorial search over 2224 ternary
systems using the \textsc{Aflow} framework~\cite{curtarolo:art65,curtarolo:art75}.  We found 102
systems that are more stable (closer to the $T=0$ K convex hull) and have a lower formation enthalpy
than the Co$_{3}$[Al$_{0.5}$, W$_{0.5}$] superalloy. All \mo{102} \cn{systems are in two-phase
  equilibrium} with the host matrix and have a relative lattice mismatch of less than or equal to
5\%.  Of these systems, \mo{37} are new---they have no reported phase
diagrams~\cite{ASMAlloyInternational,PaulingFile,Pauling}.  These new candidates, narrowed from thousands of
possibilities, suggest experimental exploration for identifying new
superalloys.  Furthermore, by eliminating  systems that are experimentally difficult to
make or contain expensive or toxic elements, we identify six particularly promising systems.

\section{Methodology}
 
\subsection{First-principles structure calculations}
 
We performed first principles calculations using the software package
\textsc{Aflow}~\cite{curtarolo:art65}.  \cn{To model an L1$_2$-based solid solution}, we used a
32-atom special quasirandom structure (SQS-32)~\cite{zunger_sqs,Jiang_JAP_2011,Saal_ActMat_2013} of
the form $X_{3}$[$A_{0.5}, B_{0.5}$] , where $X$ is one of the base elements, nickel (Ni), cobalt
(Co) or iron (Fe) (refer Fig.~\ref{ele}).  These combinations lead to 780 different ternary structures
for each base element totaling to 2340 SQS structures in 2224 different ternary systems.

All the calculations follow the \textsc{Aflow}~\cite{curtarolo:art104} standard, are hosted in the \textsc{Aflow} repository~\cite{curtarolo:art75},
and can be easily accessed by using the \mo{\textsc{RESTAPI}}~\cite{curtarolo:art92}.
Each {\it ab-initio}  calculation is performed using PAW
potentials~\cite{vasp_JPCM_1994,PAW,kresse_vasp_paw} within the  generalized gradient approximation of Perdew, Burke, and 
Ernzerhof~\cite{PBE,PBE2}, as implemented in \textsc{VASP}~\cite{kresse_vasp_1,vasp}.
The $k$-point meshes for sampling the Brillouin zone are constructed using the Monkhorst-Pack scheme~\cite{monkhorst}.
A total number of at least 10,000 $k$-points per reciprocal atom are used, and spin polarization~\cite{curtarolo:art104} is considered.
The cutoff energy is chosen to be 1.4 times the default maximum value of the three elements in the respective ternary system. 
More details are available in Ref.~\cite{curtarolo:art104}.

\begin{figure}[tb!]
\centering
\includegraphics[scale=0.25]{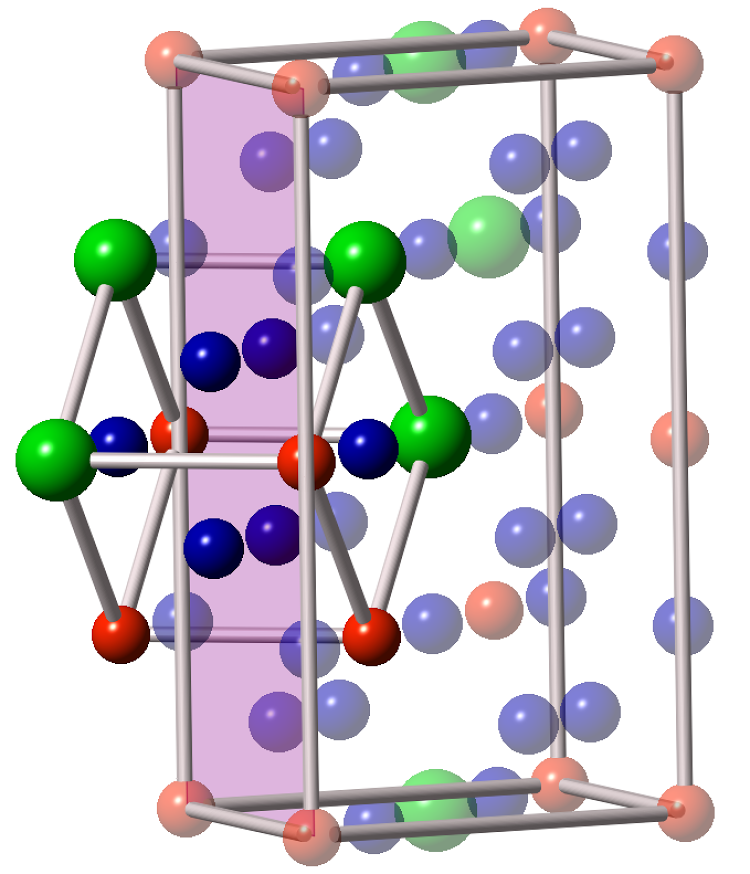}
\caption{The 32-atom special quasirandom structure (SQS-32)~\cite{Jiang_JAP_2011} used to
 \cn{ model a solid solution with an L1$_2$ structure} (smaller cube in the figure).  
The blue, red, and green atoms correspond to $X$, $A$, and $B$ in $X_{3}[A_{0.5}, B_{0.5}]$, respectively.}
\label{sqs32}
\end{figure}

The special quasirandom structure (SQS)~\cite{zunger_sqs} approach mimics the statistics of a random alloy in a small 
supercell~\cite{Jiang_ActMat_2009}.
Fig.~\ref{sqs32} depicts the 32-atom SQS~\cite{Jiang_JAP_2011} that was used for all calculations in this work. 
It is an L1$_{2}$-based structure where $X$ atoms (blue) are on the face centers of the conventional fcc cell and $A$ (red), 
$B$ (green) atoms on the corners.

\subsection{Thermodynamic property calculations}
\label{thermo}
The formation enthalpy $\left(\Delta H_{f}\right)$ is calculated for any ternary structure $X_{3}$[$A_{0.5}, B_{0.5}$] as  
%\begin{equation}
\[\Delta H_{f} = E \left(X_{3}\left[A_{0.5}, B_{0.5}\right]\right) - \sum_{m} E_{m},\]
%\end{equation}
where $E \left(X_{3}\left[A_{0.5}, B_{0.5}\right]\right)$ is the total energy per atom of the SQS-32--$X_{3}[A_{0.5}, B_{0.5}]$ structure, 
and $\sum_{m} E_{m}$ is the  sum of  \mo{total energies} of the corresponding stable, pure concentration structures.
A negative formation enthalpy characterizes a system  that
\mo{prefers an ordered} configuration over decomposition into 
its pure constituents, while unstable systems have a positive formation enthalpy. 

To approximate the phase diagram of a given alloy system, we consider the low-temperature limit in
which the behavior of the system is dictated by the ground
state~\cite{monster,monsterPGM}. %,curtarolo:mit_thesis}.
In compositional space, the set of ground state configurations defines the minimum energy surface,
also referred to as the lower-half convex hull.  All compounds above the minimum energy surface are
not stable, with the decomposition described by the hull member or facet directly below each.
The energy gained from this decomposition is geometrically represented by the distance of the
compound from the hull and quantifies the compound's tendency to decompose.  We refer to this
quantity as the decomposition energy.

While the minimum energy surface changes at finite temperature (favoring disordered structures), we
expect the $T=0$ K decomposition energy to serve as a reasonable descriptor for relative
stability. The ternary convex hulls and relevant calculations were performed\footnote{We found in
our calculations that the formation enthalpy of two compounds, namely Al$_{2}$Co and Al$_{2}$Fe
with Be$_{2}$Zn structure (the prototype numbered 549 in \textsc{Aflow}~\cite{curtarolo:art75}),
is anomalously low ($<$-1.8 eV/atom).  Similar results with this Be$_{2}$Zn structure for other
compounds were discussed previously by Taylor et al.~\cite{curtarolo:art54}.  They attribute
the erroneous results to \textsc{PAW}-pseudopotentials distributed with \textsc{VASP}.  The phase
diagrams for systems with binary combinations (Al, Co) or (Al, Fe) are generated discarding the
Be$_{2}$Zn structure in this work.}  using the phase diagram module within
\textsc{Aflow}~\cite{curtarolo:art65} (see~\hyperref[appendix:chull]{Appendix} for details).

\gh{We observe that ternary L1$_2$ phases in known superalloys have large negative formation
  enthalpies and appear near each other in Pettifor-like maps of the formation enthalpy and
  decomposition energy. Decomposition energy and formation enthalpy maps comprising all 2224 systems
  considered in this study are shown in Figs.~\ref{combined} and \ref{combined-decomp}.  All those
  systems for which decomposition energy and formation enthalpy are less than that of
  Co$_{3}$[Al$_{0.5}$,W$_{0.5}$] are included in our list of potential candidates.}

\subsection{\mo{Coherency and two-phase equilibrium with the host}}
\label{twophaseequilibrium}
\gh{Because the strain energy cost is lower, compounds with smaller
  lattice mismatch between the L1$_2$ phase and the host matrix are more likely to form coherent
  precipitates. Relative lattice mismatch ($\Delta a / a_{\mathrm{host}}$) is defined as the ratio
  of the difference between the lattice parameter of the host matrix and the precipitate
  compound, $\Delta a$, to the lattice parameter of host matrix, $a_{\mathrm{host}}$. In this work, a relative
  lattice mismatch cutoff of no more than 5\% is used to screen for potential
  superalloys.}

\begin{figure}[tb!]
\centering
\includegraphics[width=0.5\textwidth]{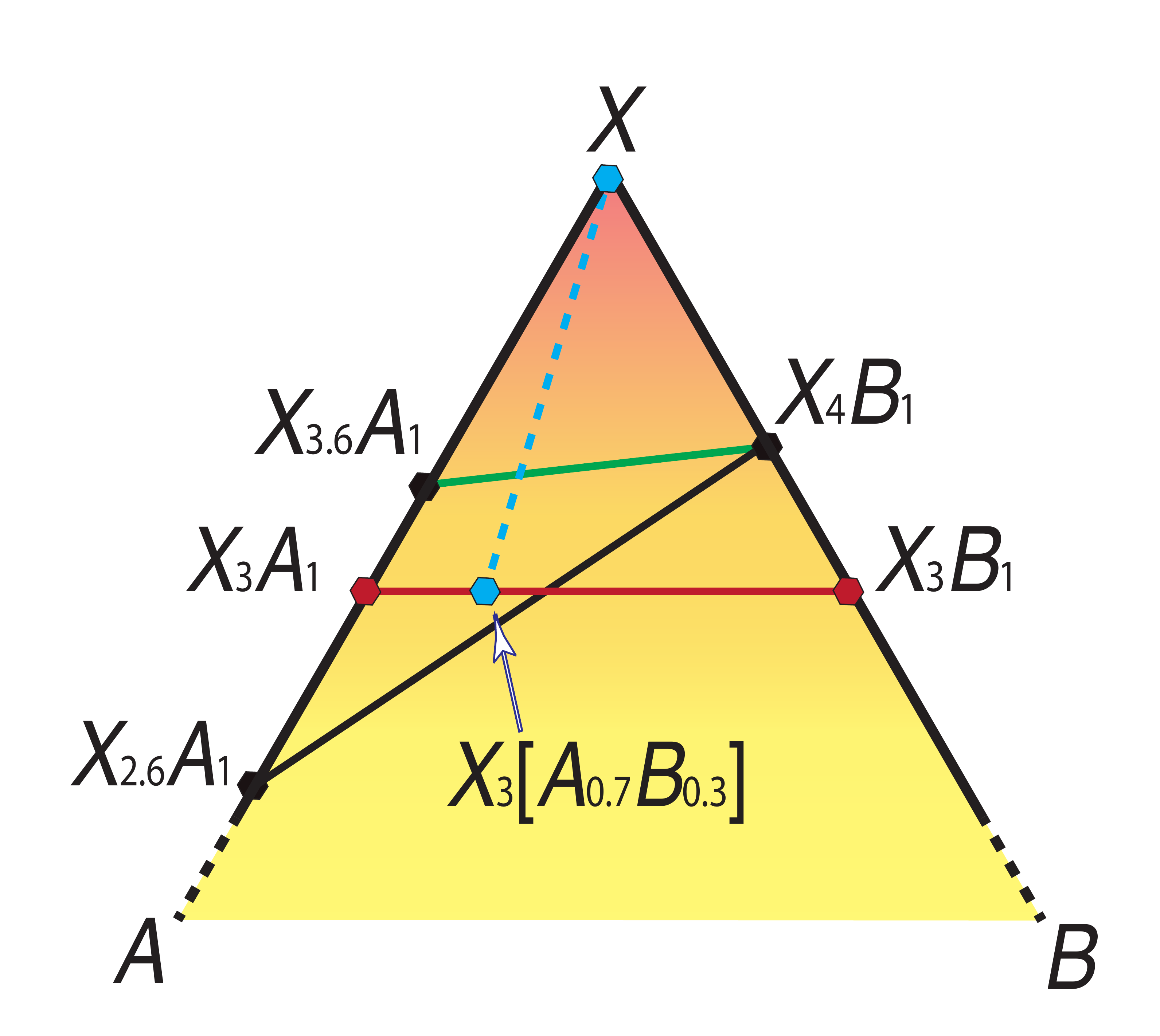}
\caption{\mo{The two-phase equilibrium
    screening criterion discussed in Sec.~\ref{twophaseequilibrium}
    (similar to Fig. 2 in Ref.~\cite{Kirklin_ActMat_2016}). If a
    tieline between the host matrix and the L1$_2$ precipitate phase
    (light blue dotted line) is intersected by the tie line for another
    phase (e.g., green line between $X_{3.6}A_1$ and $X_4B_1$) then the precipitate phase will not be in two-phase equilibrium with the
    host matrix for any concentration between $X_3A_1$ and
    $X_3B_1$. On the other hand, even if the line connecting $X_3A_1$ and $X_3B_1$ is intersected by another tie line (e.g., black line
    between $X_{2.6}A_1$ and $X_4B_1$), there may still be a
    concentration of the precipitate phase that can be in two-phase
  equilibrium with the host matrix, as show by the light-blue dotted line.}
\label{tieline}}
\end{figure}

\mo{
  %Because embedding of L1$_2$ precipitates in the host matrix is the
  %primary strengthening mechanism in conventional superalloys, we
  Because precipitate strengthening is the key mechanism for
  superalloy performance, we apply a second constraint requiring that
  the L1$_2$ precipitate phase be in two-phase equilibrium with the
  fcc host matrix.\footnote{In cases where the formation enthalpy of the SQS structure is above the convex hull,
  we project it onto the convex hull and draw the tieline between the projected point and the host
  matrix to check the two-phase equilibrium criterion.} As shown in Fig.~\ref{tieline}, this constraint is
  satisfied if a tie-line can be drawn between the host matrix (100\%
  $X$) and the L1$_2$ phase at any
  concentration($X_{3}$[$A_{x},B_{1-x}$], $0<x<1$) without intersecting
    any other tieline. We allow for this variation in the
    concentration for the minority site ($A_xB_{1-x}$) because stable
    L1$_2$ phases in experiment can vary over a wide concentration
    range~\cite{Huneau_Intermetallics_1999,Giessen_ActCrys_1965}.
  Of the 179 systems with deeper formation enthalpy and
  smaller decomposition energy than Co-Al-W, 66 systems are eliminated
  using the two-phase equilibrium criterion.}

\subsection{Bulk modulus calculations}

The bulk modulus is determined from energy-volume data calculated for
strains of $-0.02$ \textup{\AA} to $+0.02$ \textup{\AA} in steps of
$0.01$ \textup{\AA} applied to the unit cell, with at least five
calculations for each system.  The energy-volume data is fitted using
the Murnaghan equation of state~\cite{Murnaghan_PNAS_1944}.

\section{Results and analysis}

\subsection{Relative stability of SQS-32 and the distance to convex hull}

\begin{figure*}[tb!]
\centering
\includegraphics[scale=0.45]{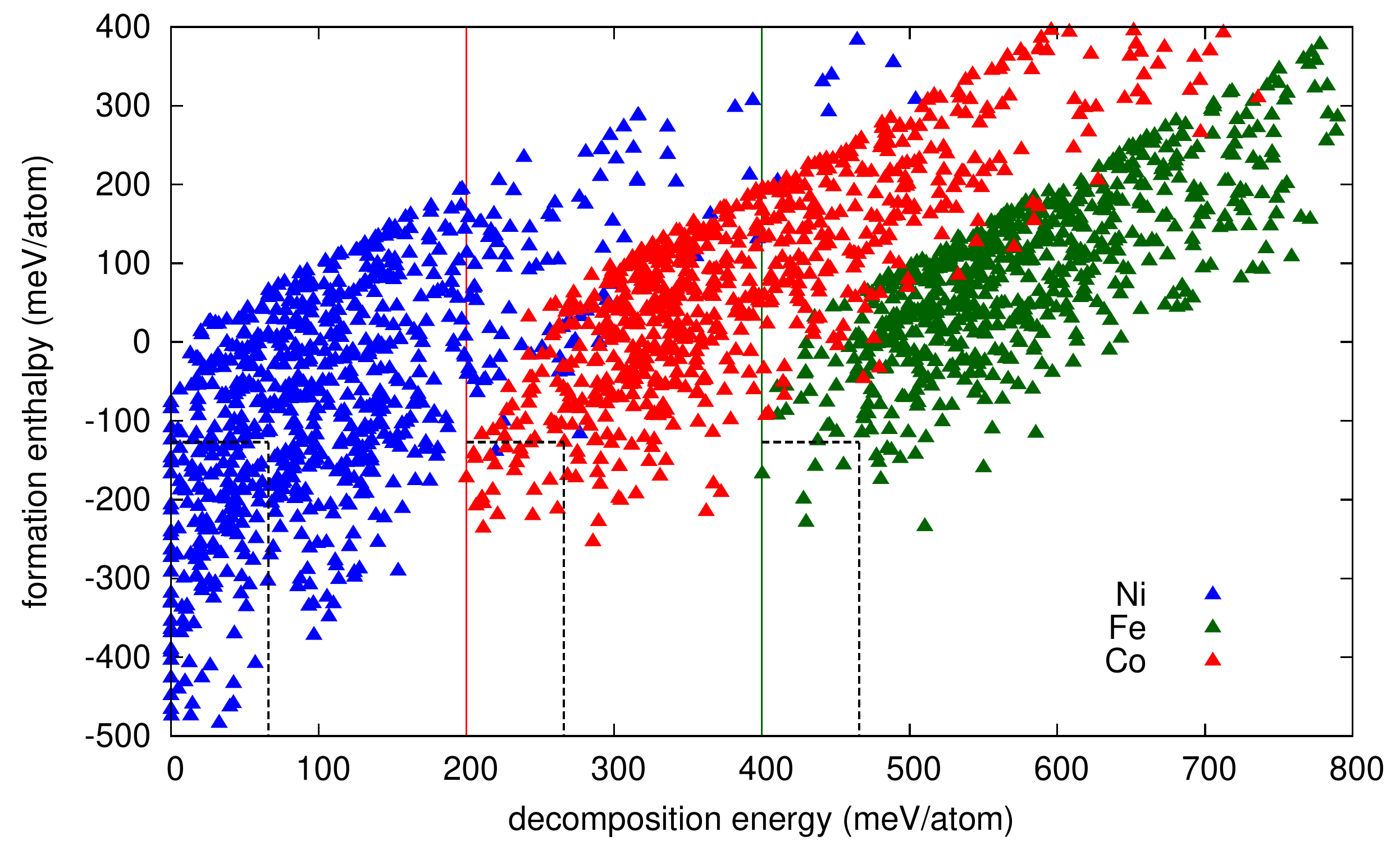}
\caption{Formation enthalpy vs. the decomposition energy for all 2224 ternary systems. 
Each triangle represents one Ni$_{3}$/Co$_{3}$/Fe$_{3}$[$A_{0.5}, B_{0.5}$] structure, where $A$ and $B$ are any two 
different elements in the periodic table from Fig.~\ref{ele}.
Co-based and Fe-based systems are displaced on the $x$-axis by 200 meV and 400 meV, respectively, for clarity. 
Ni-based, Co-based, and Fe-based systems are marked in blue, red, and green triangles, respectively. 
Systems enclosed within dotted lines are the ones identified to be better than the Co$_{3}$[Al$_{0.5}$, W$_{0.5}$] structure
with respect to these properties.}
\label{pdvse}
\end{figure*}

Fig.~\ref{pdvse} depicts the formation enthalpy $\left(\Delta H_{f}\right)$ vs. decomposition energy $\left(E_{d}\right)$ for 
all 2224 SQS-32 ternary systems with composition distinguished by color.
It is found that 2111/2224 ternary systems are compound-forming.
Each point on the plot represents one Ni$_{3}$/Co$_{3}$/Fe$_{3}$($A,B$) system, where $A$ and $B$ are any two different elements 
highlighted in Fig.~\ref{ele}. 
On average, Ni-based superalloys are thermodynamically more stable than Co- or Fe-based superalloys.

The SQS-32 structure in 179 ternary systems is found to be
thermodynamically more stable and have lower formation enthalpy than
the Co$_{3}$(Al, W) system.  These systems are enclosed within dotted
lines in Fig.~\ref{pdvse}.  Out of these systems, 152 are Ni-based, 22
are Co-based, and 5 are Fe-based.  \mo{Furthermore, 102 systems of
  these 179 are observed to be in two-phase equilibrium with
  the host matrix and have no more than  5\% relative lattice
  mismatch with respect to the respective host lattice. Of these 102 systems, 37 have no
  reported phase diagrams in \cn{standard databases}~\cite{ASMAlloyInternational,Pauling,PaulingFile}}.
Of these systems, \mo{33} are Ni-based, 3 are Co-based systems, and 1
is Fe-based.  

%The experimental phase diagrams
%complement the theoretical, first principles data available in
%\textsc{Aflow}~\cite{curtarolo:art75} with a wealth of relevant
%information}
%\footnote{One major issue in filling the gaps between theoretical and experimental materials repositories is 
%that the Wyckoff positions for many experimental structures are not reported.}.

\begin{figure*}[p]
\centering
\includegraphics[scale=0.5]{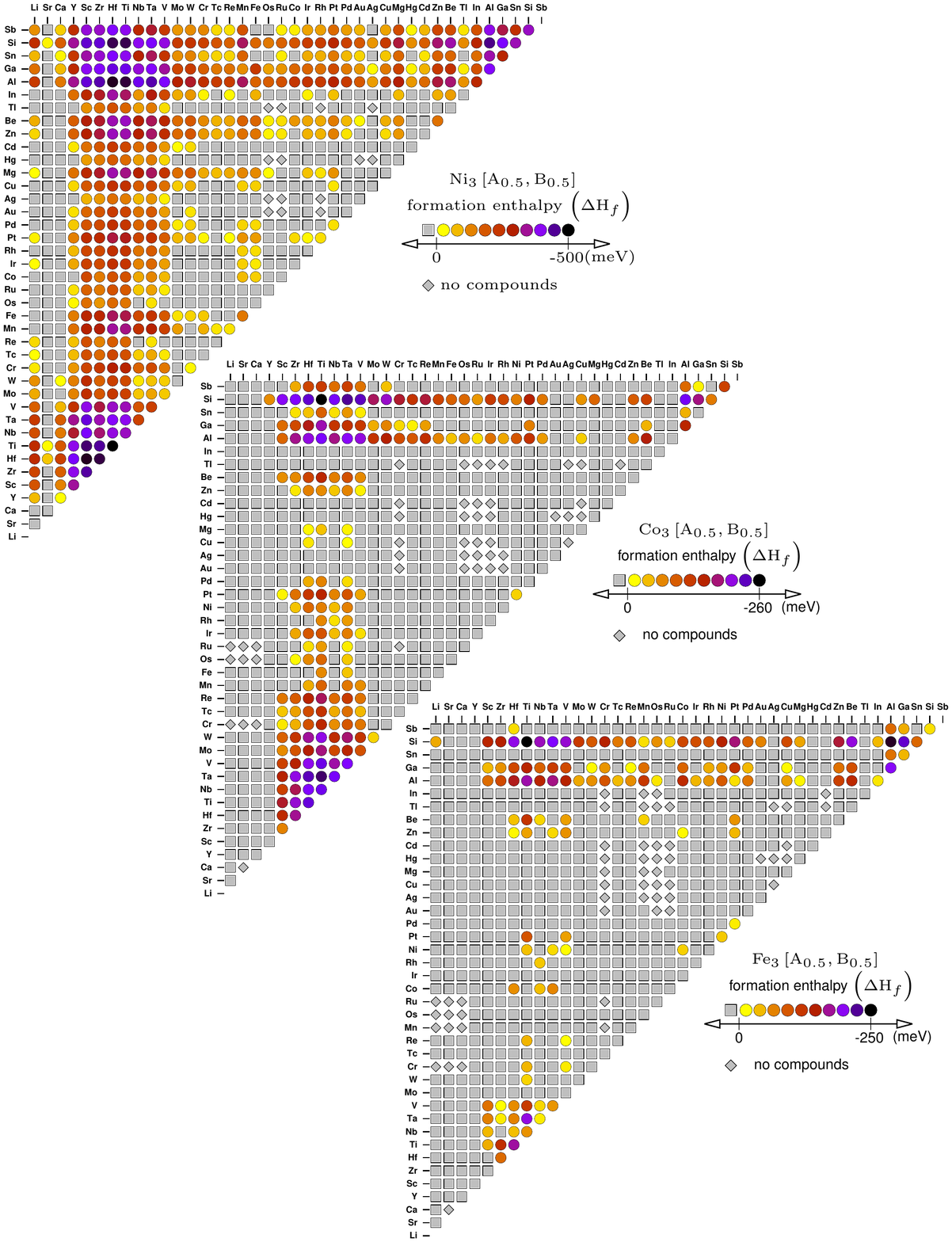}
\caption{All the elements are arranged as per the chemical scale ($\chi$) introduced by Pettifor~\cite{pettifor:1984} 
in increasing order.  
Each diamond, square, and circle represents a ternary combination $X_{3}$[$A_{0.5}, B_{0.5}$] with $X$ = Ni, Co, or Fe, and
$A,B$ specifying the elements indicated along the $x$ and $y$-axes, respectively. 
A square indicates that the SQS-32 crystal structure has a positive formation enthalpy. 
A diamond indicates that there exists no stable binary or ternary compounds in the respective ternary system. 
A colored circle indicates that the SQS-32 structure has a negative formation enthalpy.  
The color contrast from yellow to black indicates decreasing formation enthalpy of the crystal
structure in the ternary system.}
\label{combined}
\end{figure*}

\begin{figure*}[p]
\centering
\includegraphics[scale=0.5]{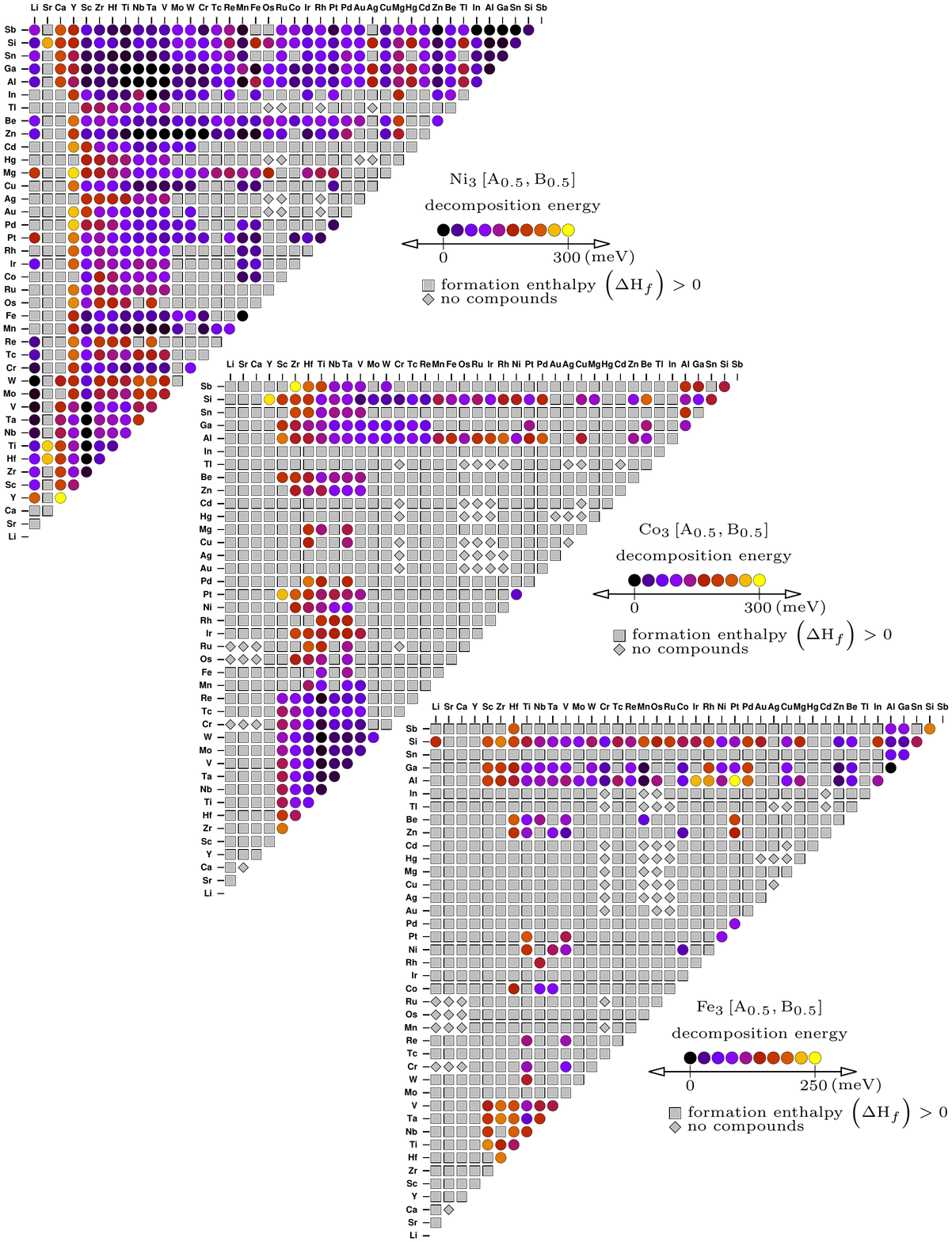}
\caption{All the elements are arranged as per the chemical scale ($\chi$) introduced by Pettifor~\cite{pettifor:1984} 
in increasing order.  
Each diamond, square, and circle  represents a ternary combination $X_{3}$[$A_{0.5}, B_{0.5}$] with $X$ = Ni, Co, or Fe, and
$A,B$ are the elements indicated along the $x$ and $y$-axes, respectively. 
A square indicates that the SQS-32 crystal structure has a positive formation enthalpy. 
A diamond indicates that there exists no stable binary or ternary compounds in the respective ternary system. 
The color contrast of the circles from yellow to black indicates increasing decomposition energy of
the crystal structure in the ternary system.}
\label{combined-decomp}
\end{figure*}

The magnitude of $\Delta H_{f}$ is closely associated with the high
temperature limit of an alloy.
If a compound has a large negative formation enthalpy, it is more likely to withstand decomposition at higher temperatures.
Fig.~\ref{pdvse} shows that many Ni-based alloys are as low as $-400$ meV compared to 
$-167$ meV of the discovered Co$_{3}$(Al, W) superalloy~\cite{Saal_ActMat_2013}.

\gh{Although the elemental form of Fe is bcc, fcc stabilizers (e.g.,
  carbon, tungsten, or nickel) can be added in small amounts to
  stabilize the fcc structure. We have modeled Fe-based
  systems with L1$_2$ precipitate-forming potential by calculating fcc
  Fe, without explicitly including the effects of the stabilizing
  additions. Had we found promising Fe systems, this rough
  approximation would have needed refinement, but all of our promising
  candidates but one turned out to be Co- or Ni-based.}

\subsection{Formation enthalpy and decomposition energy maps}
  \mo{Recognizing that ternary L1$_2$ phases in known superalloys have
    large negative formation enthalpies and small decomposition
    energies, it is useful to identify chemical trends (via the
    Pettifor chemical scale) for these two quantities. We visualize
    these trends with Pettifor-like ``formation enthalpy maps'' and
    ``decomposition energy maps'' (Figs.~\ref{combined} and
    \ref{combined-decomp}).  In the formation enthalpy maps, the formation
    enthalpy of every system computed in this work is displayed
    together, arranged in a grid ordered by the Pettifor
    scale~\cite{pettifor:1984} of the two minority components, $A,B$
    in $X_{3}$[$A_{0.5}, B_{0.5}$]. In a similar Pettifor scale grid fashion, the
    decomposition energy maps show the decomposition energy
    of every system computed in this work (Fig.~\ref{combined}). The ``islands'' of similarly
    colored compounds visible in these plots reveal
    distinct chemical trends. Many of the promising superalloy candidates
    identified in our study with no previously reported phase
    diagrams are found within these islands.}  In general, early
  $d$-block elements and $p$-block combinations dominate the list of
  favorable systems, which have both low formation enthalpies and low decomposition energies.  For Ni-based alloys,
  favorable compounds mostly comprise of transition metals Y, Sc, Zr, Hf,
  Ti, Nb, Ta, and metalloids, including Ga, Si, and Sb.  In the case
  of Co-based alloys, combinations of Zr, Hf, Ti, Nb, Ta, and Al
  define the majority of favorable compounds. Combinations of Al, Si, Hf,
  and Ti with Fe tend to produce some favorable compounds as well.  On
  the other hand, combinations with Os, Ru, and Cr tend to yield
  unstable compounds for combinations with Ni, Co, and Fe.

\subsection{Phase diagrams}
 
Ternary phase diagrams at $T=0$ K for all 2111 compound-forming systems have been plotted in this
work using the data in the open-access materials properties database
\textsc{Aflow}~\cite{aflowlib.org}.  \mo{ Convex hulls constructed from a DFT database are only as
  reliable as the database is complete. To be} \cn{robust, the database must include all possible
    structural prototypes. Our prototypes list includes essentially all known prototypes from the
    Pauling File~\cite{Pauling,PaulingFile} (a database of experimentally observed binary metallic phases) and
    binary and ternary intermetallic prototypes\footnote{Although entries with incomplete structural
      information or phases with partially occupied wyckoff positions obviously cannot be included.}
    in the ICSD~\cite{icsd1,icsd2}.  Our prototypes list also includes binary and ternary
    hypothetical structures (enumerated as in Refs.~\cite{gus_enum,enum2}). Our convex hulls were
    constructed from more than 800 DFT calculations per system. In total, 271,000 calculations were
    used for the 2111 compound-forming systems, giving us a high degree of the confidence that the
    phase stability predictions and potential superalloy candidates listed in this work are
    reasonably likely to be stable experimentally. Further evidence of the robustness of the
    calculations is given in Table~\ref{tableConv}.}% and Fig.~\ref{phasediag}.}

 \begin{table}
   \caption{\mo{Systems where the SQS structure computed in this work has a corresponding L1$_2$
       phase reported in experiment. The experimental compounds are all close to the stoichiometry
       of the SQS structure, $X_{24}[A_{4},B_{4}]$.}}
 \centering
  \label{tableConv}
  \begin{tabular}{ll}

\hline

SQS  & Exp.  \\

\hline 

Al$_{0.5}$Cr$_{0.5}$Ni$_{3}$  & Al$_{0.8}$Cr$_{0.2}$Ni$_{3}$ \cite{UlHaq_JMMM_1986}\\
Al$_{0.5}$Cu$_{0.5}$Ni$_{3}$  & Al$_{1}$Cu$_{0.28}$Ni$_{2.72}$ \cite{Mishima_ActaMatt_1985} \\
Al$_{0.5}$Ga$_{0.5}$Ni$_{3}$  & Al$_{0.5}$Ga$_{0.5}$Ni$_{3}$ \cite{ochiai1984lattice}\\
Al$_{0.5}$Hf$_{0.5}$Ni$_{3}$  & Al$_{0.99}$Hf$_{0.01}$Ni$_{3}$ \cite{rao1992effect}\\
Al$_{0.5}$Nb$_{0.5}$Ni$_{3}$ & Al$_{0.65}$Nb$_{0.35}$Ni$_{3}$  \cite{Mints_RJIC_1962} \\
Al$_{0.5}$Ni$_{3}$Pt$_{0.5}$ & Al$_{1}$Ni$_{2.48}$Pt$_{0.52}$  \cite{Mishima_ActaMatt_1985}  \\
Al$_{0.5}$Ni$_{3}$Si$_{0.5}$ & Al$_{0.6}$Ni$_{3}$Si$_{0.4}$  \cite{Mishima_ActaMatt_1985}  \\
Al$_{0.5}$Ni$_{3}$Sn$_{0.5}$ & Al$_{0.8}$Ni$_{3}$Sn$_{0.2}$  \cite{UlHaq_JMMM_1986}  \\ 
Al$_{0.5}$Ni$_{3}$Ta$_{0.5}$ & Al$_{0.76}$Ni$_{3}$Ta$_{0.24}$ \cite{Giessen_ActCrys_1965}  \\ 
Al$_{0.5}$Ni$_{3}$Ti$_{0.5}$ & Al$_{1}$Ni$_{2.8}$Ti$_{0.2}$ \cite{Huneau_Intermetallics_1999}\\ 
Al$_{0.5}$Ni$_{3}$V$_{0.5}$ &Al$_{0.28}$Ni$_{3}$V$_{0.2}$\cite{UlHaq_JMMM_1986} \\ 
Co$_{3}$Ti$_{0.5}$V$_{0.5}$ & Co$_{3}$Ti$_{0.87}$V$_{0.13}$ \cite{liu1986alloying}\\
Ga$_{0.5}$Hf$_4$Ni$_{3}$ & Ga$_{0.88}$Hf$_{0.12}$Ni$_{3}$\cite{Mishima_ActaMatt_1985}   \\
Ga$_{0.5}$Nb$_4$Ni$_{3}$ &Ga$_{0.84}$Nb$_{0.16}$Ni$_{3}$ \cite{Mishima_ActaMatt_1985}  \\ 
Ga$_{0.5}$Ni$_{3}$Sb$_{0.5}$ &Ga$_{0.92}$Ni$_{3}$Sb$_{0.08}$\cite{Mishima_ActaMatt_1985}  \\ 
Ga$_{0.5}$Ni$_{3}$Si$_{0.5}$ &Ga$_{0.4}$Ni$_{3}$Si$_{0.6}$\cite{Mishima_ActaMatt_1985}  \\ 
Ga$_{0.5}$Ni$_{3}$Sn$_{0.5}$ &Ga$_{0.84}$Ni$_{3}$Sn$_{0.16}$\cite{Mishima_ActaMatt_1985}  \\ 
Ga$_{0.5}$Ni$_{3}$Ta$_{0.5}$ &Ga$_{0.68}$Ni$_{3}$Ta$_{0.32}$\cite{Mishima_ActaMatt_1985} \\
Ga$_{0.5}$Ni$_{3}$Ti$_{0.5}$ & Ga$_{0.84}$Ni$_{3}$Ti$_{0.16}$ \cite{Mishima_ActaMatt_1985}\\
Ga$_{0.5}$Ni$_{3}$V$_{0.5}$ & Ga$_{0.76}$Ni$_{3}$V$_{0.24}$\cite{Mishima_ActaMatt_1985} \\

\hline
\hline
\end{tabular}
\end{table}

%\begin{figure*}[tb]
%\centering
%\includegraphics[scale=0.35]{fig7}
%\caption{Comparison of experimentally reported and computational predicted phases for 
%the Al-Ni-Ta and Al-Ni-Ti systems.
%Phases highlighted in red are exclusively experimentally reported (in the ASM Database), while phases
%highlighted in blue are exclusively computationally reported (in \textsc{Aflow}).
%Phases in green indicate a match between experimental and predicted results.
%%    Al-Ni-Ta.  The green dot
%%  indicates that the compound is reported in
%%  both \textsc{Aflow} and ASM databases. The blue dots
%%  indicates that the compound is reported only in \textsc{Aflow}
%%  database. The red dot indicates that the compound is reported
%%  in ASM database. \cn{I'll add one more phase diagram and description
%%    later.}  
%}
%\label{phasediag}
%\end{figure*}

\mo{The ternary phase diagrams of all 2111 compound-forming systems
are included in the Supplementary Material accompanying this
work
\co{and are available online via \href{http://aflow.org/superalloys}{http://aflow.org/superalloys}.}
They were created with the phase diagram module within
\textsc{Aflow}.  In almost all cases, the \textsc{Aflow} convex hulls
contain more phases than reported in the experimental databases. In
some cases, this may indicate an opportunity for further
experimental study, but it is likely that \cn{some} of these DFT ground
states are low temperature phases and \cn{are therefore} kinetically
inaccessible, which explains why they are not reported in experimental
phase diagrams. }

%\mo{As typical examples, we examine the convex hulls for Al-Ni-Ta \cn{and Al-Ni-Ti} in
%  Fig.~\ref{phasediag}. \cn{In these two systems, the SQS structure} appears on the convex hull and
%  there are experimentally known L1$_2$ precipitate phases at finite temperatures. The figure
%  compares the structures from the \textsc{Aflow} convex hull to all experimentally reported phases (except
%  those for which structural information [prototype, wycoff positions, etc.] is not given). The agreement is
%  good---in most cases where the experiment and computation do not match exactly, the
%  computational phases are ordered realizations of experimentally disordered phases. Essentially,
%  there are no disagreements, merely differences that can be accounted for by finite temperature
%  effects.}

\cn{There are 66 systems which meet all
  our criteria discussed in Secs.~\ref{thermo} and \ref{twophaseequilibrium} and for which there are
  published phase diagrams.  In 20 of those systems, the predicted L1$_2$ phase is validated by an
  experimentally reported L1$_2$ phase at nearby concentrations. In 37 cases, the phase diagrams are
  incompete in the region of interest. In the remaining eight cases, three have fcc solid
  solutions near our composition, three report disordered $\chi$-like phases or unknown structures,
  one has a disordered D0$_{24}$ structure (closely related to L1$_2$ and a precipitate phase in
  some superalloys), and one reports the structure prototype Mg$_6$Cu$_{16}$Si$_7$.}
\subsection{Density of superalloys}

\begin{figure}[tb!]
%\hspace*{-2cm}
\centering
\includegraphics[width=\linewidth]{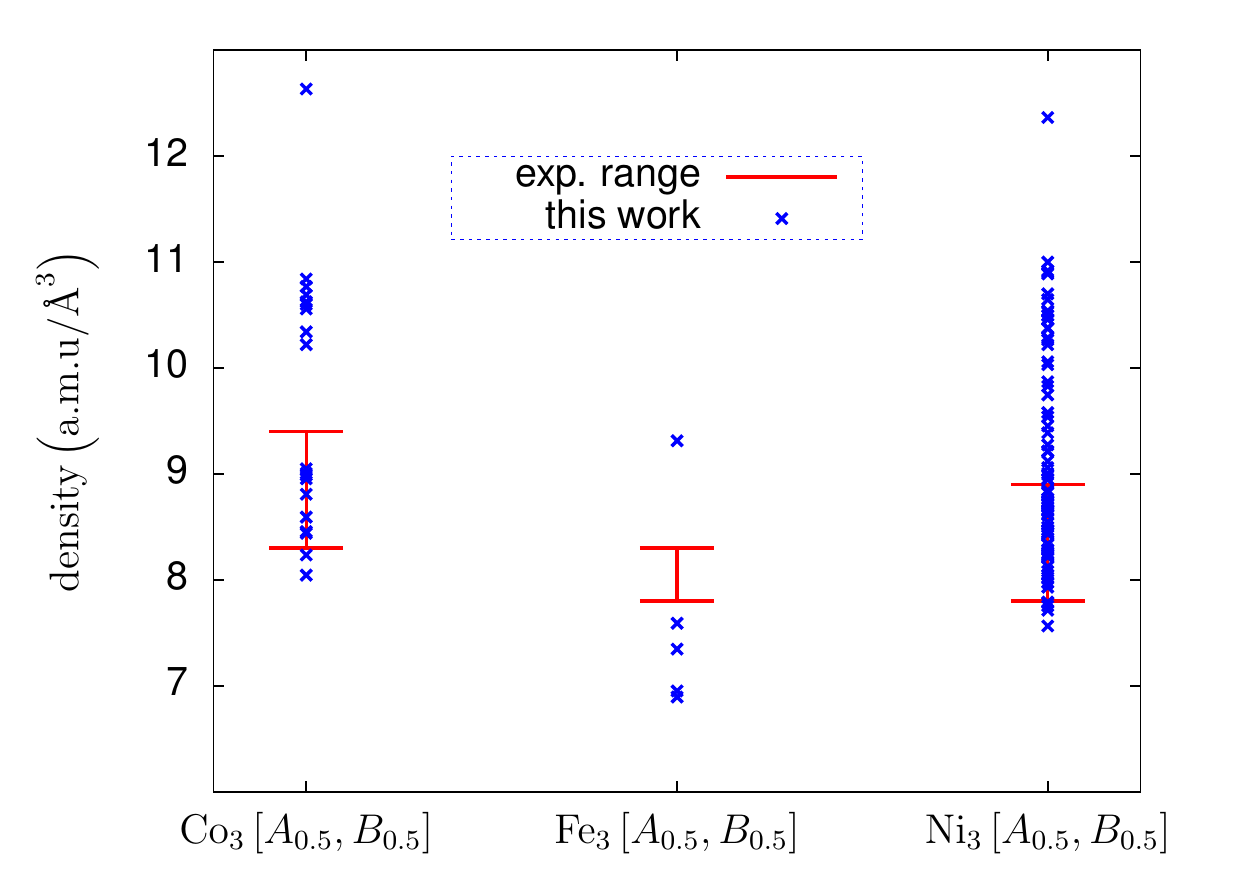}
\caption{A comparison between the density range for the theoretical calculations performed in this work and modern superalloys. Densities are computed for \mo{102} ternary systems screened from the 2224 systems computed in this work. The red line shows the range of density for commercially-available superalloys at present.}
\label{densityPlot}
\end{figure}

Low density and high-temperature strength are two critical properties of superalloys for any application.  
For example, increased density can result in higher stress on mating components in aircraft gas turbines~\cite{Donachie_ASM_2002}.
A comparison between the density range for theoretical calculations performed in this work and modern superalloys is listed 
in Fig.~\ref{densityPlot}. 
Of the theoretical ternary combinations, there are 5 Ni-based alloys 2 Co-based and 4 Fe-based alloys with density less than the range of commercially-available 
superalloys.
This certainly warrants further analysis of mechanical properties of these alloys, which may yield novel lightweight, high-strength superalloys.
 
\begin{figure}[tb!]
\begin{center}
\includegraphics[width=\linewidth]{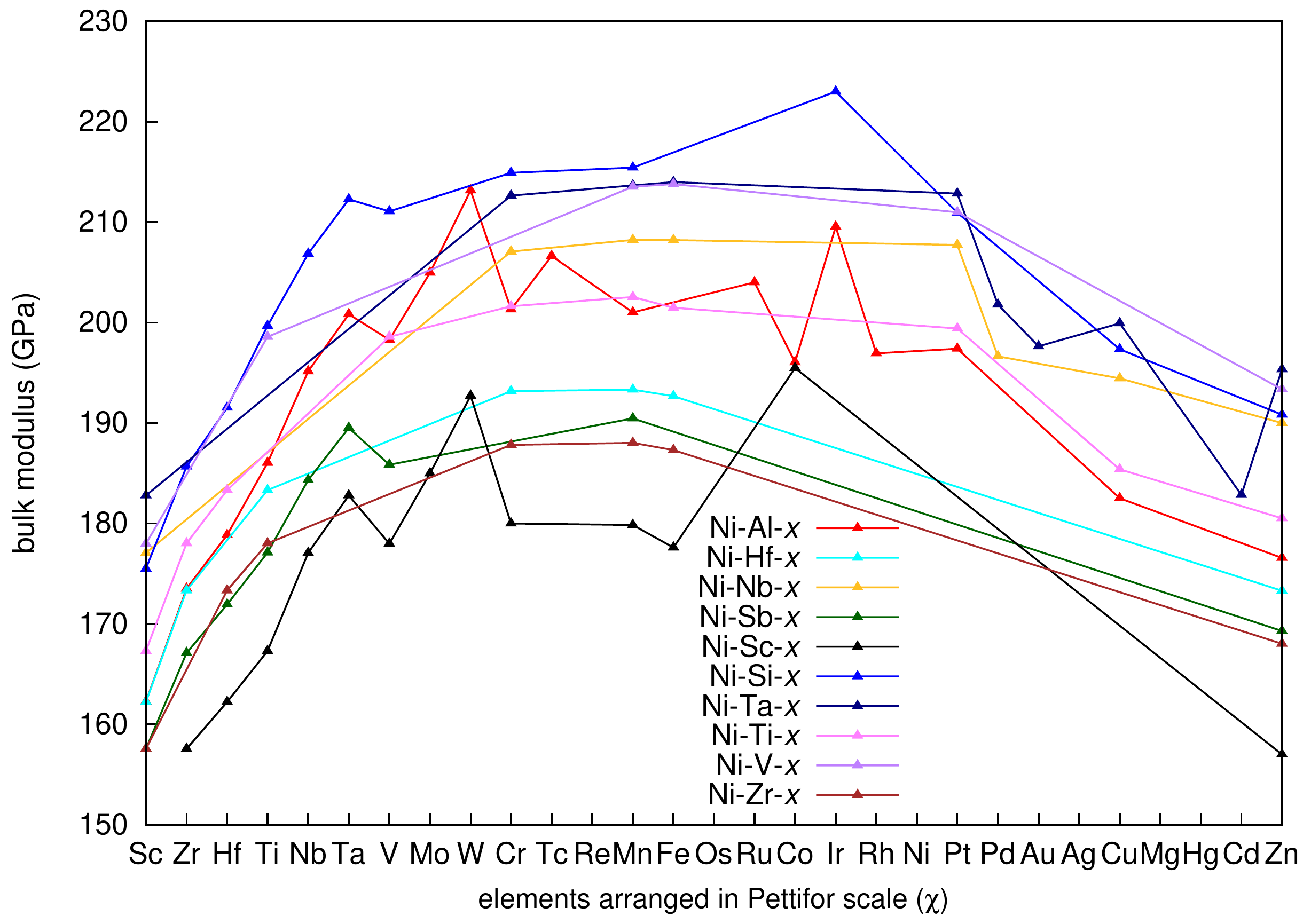}
\end{center}
\caption{The magnitude of the bulk modulus for Ni-$A$-$x$ 
($A$ = Al, Hf, Nb, Sb, Sc, Si, Ta, Ti, V, and Zr) systems with the $x$-axis arranged according
to the $\chi$ scale in Pettifor maps.
In general, the systems display a maximum in the bulk modulus at or before Ni.
Only systems with simultaneously lower $E_{d}$ and $\Delta H_{f}$ than Co$_{3}$[Al$_{0.5}$,W$_{0.5}$] are plotted.}
\label{bulkni}
\end{figure}

\subsection{Bulk modulus}

For the aforementioned systems with simultaneously lower $E_{d}$ and $\Delta H_{f}$ than 
the Co$_{3}$(Al, W) system, the bulk modulus is computed in this work.
All the Co- and Fe-based alloys have a bulk modulus of at least 200 GPa.
This is consistent with the observation that commercial Co-based alloys have better mechanical properties than many 
Ni-based alloys~\cite{Sato_Science_2006}.

\begin{figure}[tb!]
\centering
\includegraphics[width=\linewidth]{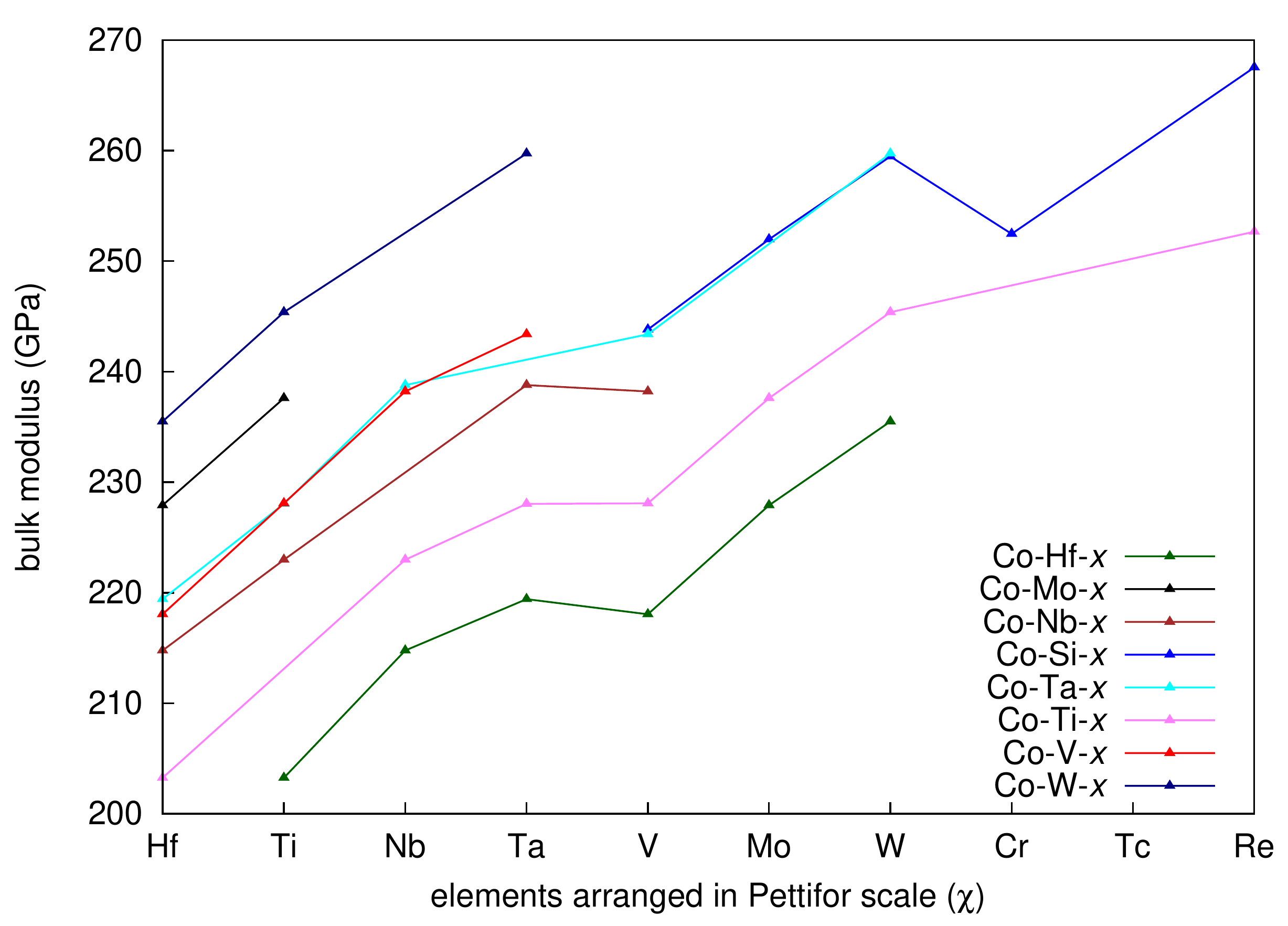}

\caption{The magnitude of the bulk modulus for Co-$A$-$x$ ($A$ = Hf, Mo, Nb, Si, Ta, Ti, V, and W) 
systems with the $x$-axis arranged according to the $\chi$ scale in Pettifor maps.
In general, the magnitude of the bulk modulus increases with $\chi$ up to Re.
Only systems with simultaneously lower $E_{d}$ and $\Delta H_{f}$ than Co$_{3}$[Al$_{0.5}$, W$_{0.5}$] are plotted.}
\label{bulkco}
\end{figure}

Figs.~\ref{bulkni} and \ref{bulkco} depict the magnitude of the bulk modulus for Ni-$A$-$x$ ($A$ = Al, Hf, Nb, Sb, Sc, Si, Ta, Ti, V, and Zr) 
and Co-$A$-$x$ ($A$ = Hf, Mo, Nb, Si, Ta, Ti, V, and W) systems. 
$x$ is the third element in the ternary system and arranged along the $x$-axis of the plot in increasing order of the 
Pettifor chemical scale ($\chi$).  
The bulk modulus of ternary alloys of the form Ni-$A$-$x$ 
reaches a maximum at or before Ni.
In case of Co-$A$-$x$ systems, the bulk modulus increases with increasing $\chi$ up to Re.

The magnitude of the bulk modulus suggests that Co-based superalloys are
particularly resistant to compression compared to Ni-based superalloys.
68 ternary systems with simultaneously lower $E_{d}$ and $\Delta H_{f}$ than Co$_{3}$[Al$_{0.5}$,W$_{0.5}$] have 
bulk moduli greater than 200 GPa.

\subsection{Promising Candidates}
\cn{Table~\ref{promising} lists the 37 systems that are predicted to have stable precipitate-forming
  L1$_2$ phases and for which there are no reported phase diagrams in standard databases~\cite{ASMAlloyInternational,PaulingFile,Pauling}.
\cn{Avoiding elements (i.e., Au, Be, Cd, Ga, Hg, Ir, In, Li, Os, Pd, Pt, Re, Rh, Ru, Sb, Sc, Tc, and Tl),
that are toxic, expensive, or have low melting temperatures (which can result in difficulty incorporating them in alloy synthesis),}
we prioritize this list into a smaller set of six candidate superalloy systems. These are
denoted by boxes in Table~\ref{promising}.}

\begin{table*}[th!]
  \caption{\mo{Candidates for precipitate-forming systems that have no previously reported phase
      diagrams in standard databases~\cite{Pauling,PaulingFile,icsd1,icsd2}. These have a smaller
      decomposition energy and a lower formation enthalpy than the Co$_{3}$(Al, W) superalloy. All
      are in stable two-phase equilibrium with the host matrix and have a relative lattice mismatch
      with the host matrix of less than or equal to 5\%.  Promising candidates (see section 3.6) are
      boxed. `$\ast \ast \ast$' indicates that the quantity is not computed in this work.}}
\label{promising}
%\begin{ruledtabular}
\centering
\begin{tabular}{llllll}
\hline 
\centering
System						&	Formation	&	Decomposition	&	Density			&	Bulk		& Relative lattice\\
							&	enthalpy	&	energy			&	[gm/cm$^{3}$]	&	modulus	& mismatch	\\
							&	[meV]		&	[meV/atom]		&					&	[GPa]	& [$\%$]	\\
\hline
Al$_{4}$Ni$_{24}$Rh$_{4}$	& 	$-$189	& 	49	& 	8.71	& 	197	& 	$-$2	\\
Au$_{4}$Ni$_{24}$Ta$_{4}$& $-$142 & 46 & 12.17 & 198 & $-$5 \\
Be$_{4}$Fe$_{4}$Ni$_{24}$  & $-$129 & 40  & 8.20 & 206	&1\\
Be$_{4}$Ga$_{4}$Ni$_{24}$	& 	$-$203	& 	59	& 	8.33	& 	184	& 	0	\\
Be$_{4}$Mn$_{4}$Ni$_{24}$	& 	$-$132	& 	43	& 	8.12	& 	***	& 	1	\\
Be$_{4}$Nb$_{4}$Ni$_{24}$ & $-$237 & 37 & 8.38 & 198 & $-$1\\
Be$_{4}$Ni$_{24}$Sb$_{4}$	& 	$-$159	& 	59	& 	8.71	& 	177	& 	$-$2	\\
Be$_{4}$Ni$_{24}$Si$_{4}$	& 	$-$298	& 	48	& 	7.78	& 	201	& 	1	\\
Be$_{4}$Ni$_{24}$Ta$_{4}$ & $-$269 & 33 & 10.02 & 204 & $-$1\\
Be$_{4}$Ni$_{24}$Ti$_{4}$	& 	$-$308	& 	53	& 	7.79	& 	189	& 	0	\\
Be$_{4}$Ni$_{24}$V$_{4}$ & $-$225 & 21 & 8.07 & 203	&1\\
Be$_{4}$Ni$_{24}$W$_{4}$ & $-$144 & 44 & 10.23 & 219 &$-$1\\
%Cd$_{4}$Ni$_{24}$Ta$_{4}$ & $-$154 & 35& 10.62 & 183 & $-$5\\
\framebox{Co$_{24}$Nb$_{4}$V$_{4}$}	& 	$-$156	& 	19	& 	9.05	& 	238	& 	$-$2	\\
Co$_{4}$Ni$_{24}$Sc$_{4}$	& 	$-$166	& 	55	& 	8.04	& 	169	& 	$-$3	\\
Co$_{24}$Re$_{4}$Ti$_{4}$	& 	$-$142	& 	5	& 	10.69	& 	253	& 	$-$2	\\
\framebox{Co$_{24}$Ta$_{4}$V$_{4}$}	& 	$-$189	& 	18	& 	10.62	& 	243	& 	$-$2	\\
Fe$_{24}$Ga$_{4}$Si$_{4}$	& 	$-$200	& 	28	& 	7.59	& 	***	& 	$-$4	\\
Ga$_{4}$Ir$_{4}$Ni$_{24}$	& 	$-$129	& 	27	& 	11.00	& 	209	& 	$-$2	\\
%Hf$_{4}$Ni$_{24}$Sb$_{4}$	& 	$-$358	& 	16	& 	10.50	& 	172	& 	$-$6	\\
\framebox{Hf$_{4}$Ni$_{24}$Si$_{4}$}	& 	$-$459	& 	42	& 	9.83	& 	192	& 	$-$3	\\
%In$_{4}$Ni$_{24}$Ta$_{4}$	& 	$-$246	& 	0	& 	10.65	& 	186	& 	$-$5	\\
In$_{4}$Ni$_{24}$V$_{4}$ & $-$165 & 14 & 8.91 & 182 & $-$4 \\
%In$_{4}$Ni$_{24}$Zr$_{4}$	& 	$-$253	& 	39	& 	8.84	& 	182	& 	$-$6	\\
Ir$_{4}$Ni$_{24}$Si$_{4}$	& 	$-$184	& 	55	& 	10.54	& 	223	& 	$-$1	\\
\framebox{Mn$_{4}$Ni$_{24}$Sb$_{4}$}	& 	$-$151	& 	8	& 	9.06	& 	184	& 	$-$4	\\
Nb$_{4}$Ni$_{24}$Pd$_{4}$	& 	$-$129	& 	52	& 	9.39	& 	197	& 	$-$4	\\
Nb$_{4}$Ni$_{24}$Pt$_{4}$	& 	$-$172	& 	48	& 	10.89	& 	208	& 	$-$4	\\
Nb$_{4}$Ni$_{24}$Zn$_{4}$	& 	$-$241	& 	0	& 	8.95	& 	190	& 	$-$3	\\
Ni$_{24}$Pd$_{4}$Ta$_{4}$	& 	$-$160	& 	51	& 	10.92	& 	202	& 	$-$4	\\
Ni$_{24}$Pt$_{4}$Si$_{4}$	& 	$-$228	& 	39	& 	10.46	& 	211	& 	$-$2	\\
Ni$_{24}$Pt$_{4}$Ta$_{4}$	& 	$-$202	& 	45	& 	12.36	& 	213	& 	$-$4	\\
Ni$_{24}$Pt$_{4}$Ti$_{4}$	& 	$-$250	& 	58	& 	10.38	& 	199	& 	$-$3	\\
%Ni$_{24}$Sb$_{4}$Sc$_{4}$	& 	$-$320	& 	47	& 	8.43	& 	158	& 	$-$6	\\
\framebox{Ni$_{24}$Sb$_{4}$Si$_{4}$}	& 	$-$310	& 	21	& 	8.82	& 	187	& 	$-$3	\\
\framebox{Ni$_{24}$Sb$_{4}$Ti$_{4}$}	& 	$-$335	& 	11	& 	8.72	& 	177	& 	$-$5	\\
Ni$_{24}$Sc$_{4}$Zn$_{4}$	& 	$-$241	& 	39	& 	7.97	& 	157	& 	$-$4	\\
Ni$_{24}$Si$_{4}$Sn$_{4}$	& 	$-$303	& 	26	& 	8.76	& 	185	& 	$-$3	\\
Ni$_{24}$Ta$_{4}$Zn$_{4}$	& 	$-$274	& 	0	& 	10.49	& 	195	& 	$-$3	\\
Ni$_{24}$V$_{4}$Zn$_{4}$	& 	$-$213	& 	0	& 	8.66	& 	193	& 	$-$1	\\
Ni$_{24}$W$_{4}$Zn$_{4}$	& 	$-$147	& 	0	& 	10.70	& 	210	& 	$-$2	\\
Ni$_{24}$Zn$_{4}$Zr$_{4}$	& 	$-$261	& 	48	& 	8.61	& 	168	& 	$-$4	\\
\hline
\vspace{1cm}
%\hline
\end{tabular}
%\end{ruledtabular}
\end{table*}

\section{Conclusion}

\cn{We used DFT calculations to search for new ternary systems with L1$_{2}$ precipitate-forming potential. We examined} a total of 2224 different ternary systems comprising 41 different elements. 
The Pettifor-type formation enthalpy and decomposition energy maps
(Fig.~\ref{combined} and \ref{combined-decomp}) introduced in this work reveal that combinations of early $d$-block and $p$-block 
elements tend to form stable superalloy systems with base-elements Ni, Co, and Fe. 
Ni-based superalloys tend to be thermodynamically more stable than Co- or Fe-based superalloys.  

A total of \mo{102} ternary systems are found to have lower formation enthalpy  and decomposition energy than the recently discovered 
Co$_{3}$[Al$_{0.5}$,W$_{0.5}$] superalloy. All the systems are
observed to be in two-phase equilibrium with the host matrix and have
a lattice mismatch of less than \mo{or equal to} 5\% with the host matrix.
Further analysis should be done for these systems with, e.g., cluster expansion~\cite{sanchez_ducastelle_gratias:psca_1984_ce,deFontaine_ssp_1994,Zunger_NATO_1994}
in the interest of experimental verification.
Of these, 37 systems have no experimental phase diagram reported in literature. A comparison between the density range for our theoretical systems and modern superalloys reveal many candidate low-density superalloys. 
Co-based superalloys are observed to have a higher bulk modulus than Ni- and Fe-based alloys.  
Based on cost, experimental difficulty, and toxicity, we prioritize a shorter list of six
promising superalloy systems (see Table 2).

\section{Acknowledgments}
The authors thank Eric Perim, Eric Gossett, M. Buongiorno Nardelli, M. Fornari, S. Butenko, and C. Toher for useful discussion.
Funding from ONR (MURI N00014-13-1-0635).
C. Oses acknowledges support from the National Science Foundation Graduate Research Fellowship under Grant No. DGF1106401.
Calculations were performed at the \emph{Duke University Center for Materials Genomics} and at the \emph{BYU Fulton Supercomputing Lab}.

%\section{Appendix}
\appendix

\section{Phase Diagram (Convex Hull) Analysis}
\label{appendix:chull}

\begin{figure*}
\centering
\begin{subfigure}
\centering
\includegraphics[width=0.45\textwidth]{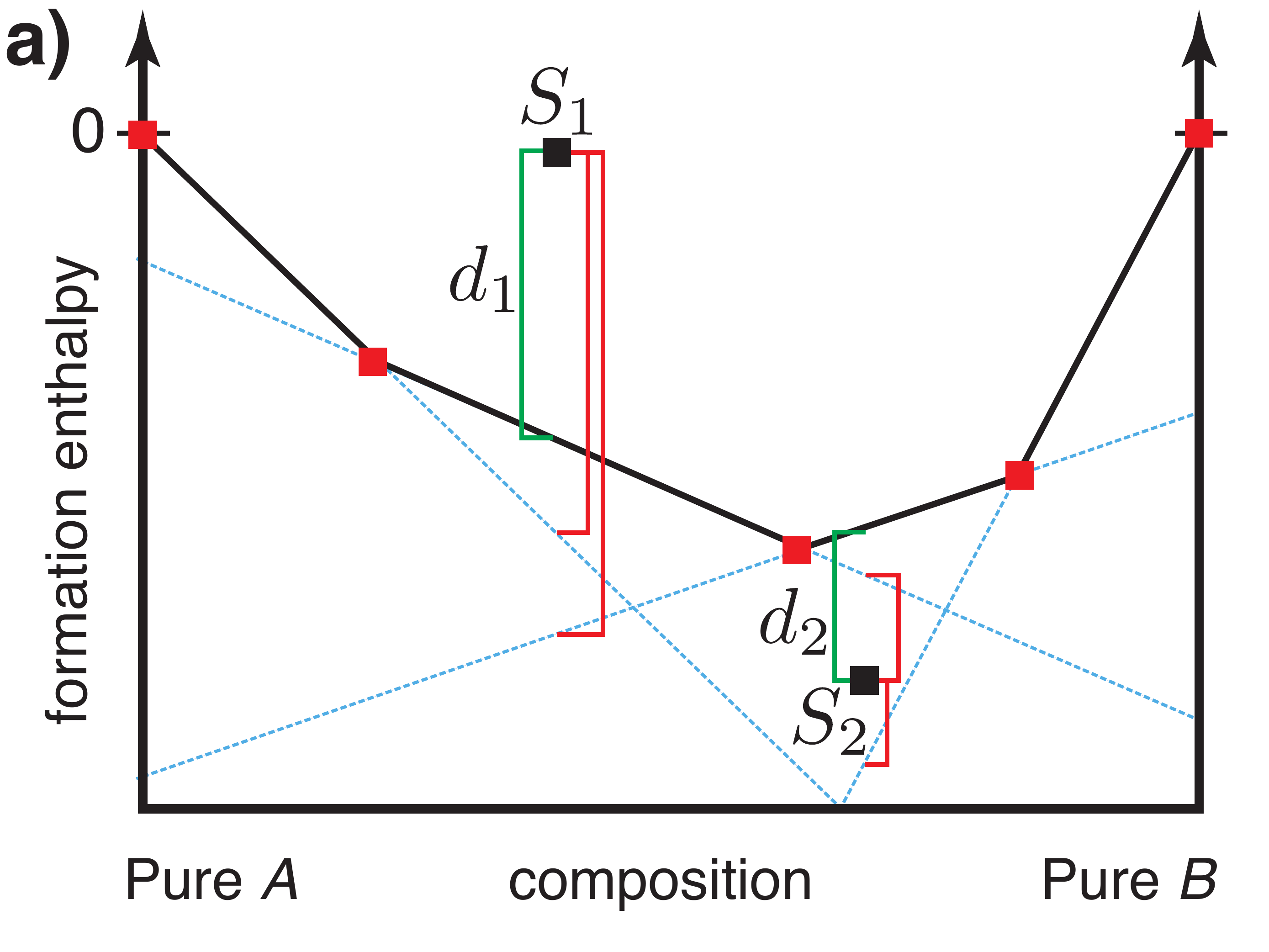}
\end{subfigure}
~
\begin{subfigure}
\centering
\includegraphics[width=0.45\textwidth]{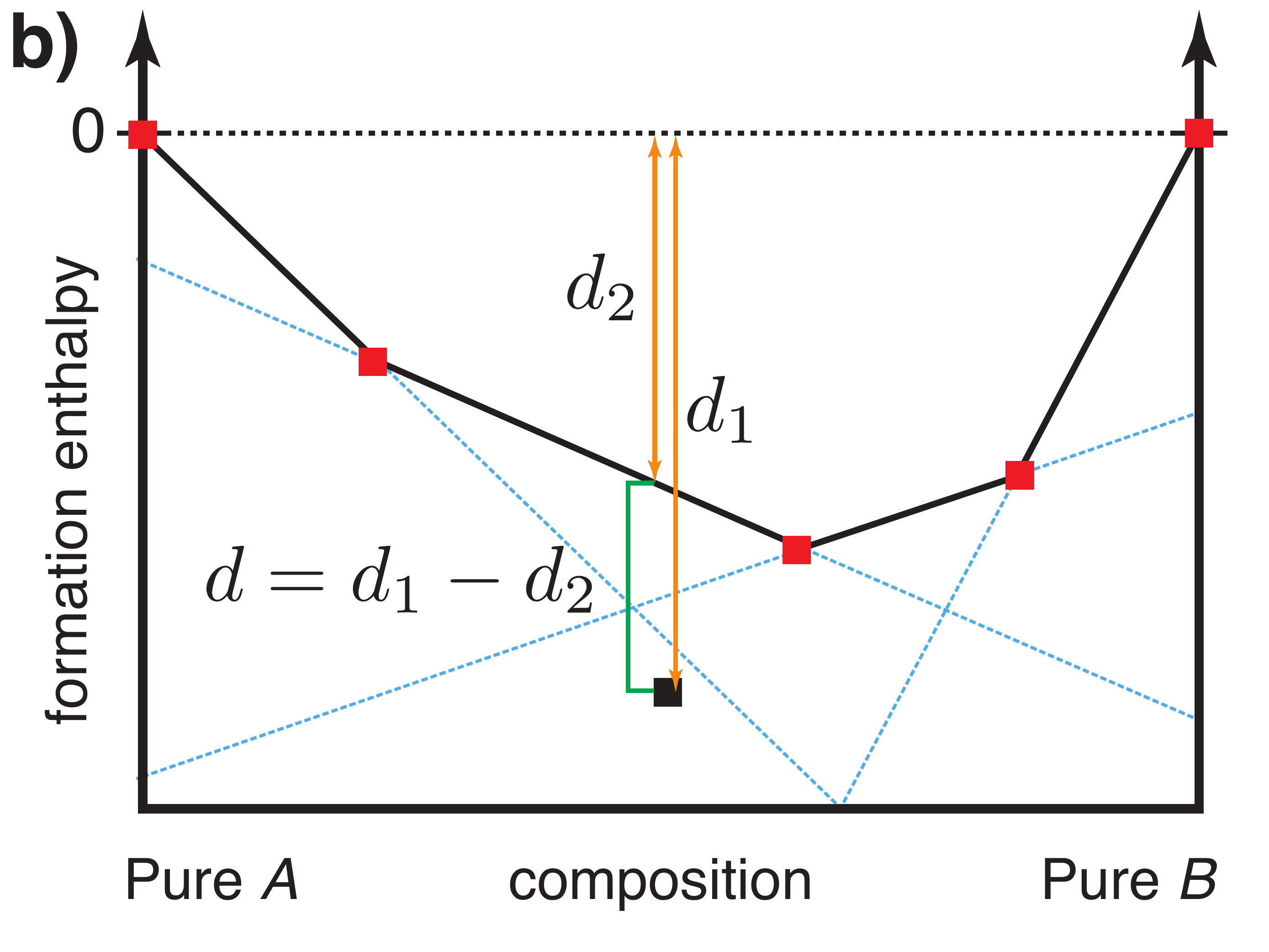}
\end{subfigure}
\caption{
Distance to the $T=0$ K convex hull algorithm.
\textbf{a)} The correct distance (shown in green) for $d_1$ is the minimum distance of structure $S_1$ to all hyperplanes defining the convex hull. 
In case of structure $S_2$, the minimum distance is not $d_2$ (green line), an artifact of the hyperplane 
description for hull facets.
\textbf{b)} Projecting the points to the zero energy line guarantees that all points will lie within the hull, 
thus enabling the use of minimization algorithm to calculate the correct distance. 
The distance to the hull $d$ is given as the difference of the projected distance $d_2$ from the distance to the zero energy line $d_1$.
}
\label{convexHull}
\end{figure*}

We construct the $T=0$ K convex hull using the phase diagram module within
\textsc{Aflow}~\cite{curtarolo:art65}.
Elements of this implementation were inspired by the \textsc{qhull} algorithm~\cite{qhull}.
For $k$-nary systems, it computes the distances to the hull with the following considerations.
Let a facet of the convex hull, i.e., a hyperplane, be described by, 
\begin{equation}
a_{0} + \sum_{m=1}^{k} a_{m}x_{m} = 0.
\end{equation}
Here $a_{1},\ldots,a_{k}$ define the normals of the hyperplane while the constant $a_{0}$ uniquely defines the hyperplane in space. 
This is a simple extension of the familiar 3-D plane equation $Ax + By+ Cz = D$.
Let a $k$-nary structure have the coordinates denoted in $k$-dimensional 
space as $c_{1},c_{2},\ldots,c_{k}$, where $c_{1},\ldots,c_{k-1}$, are the concentrations of the $k-1$ elements in a $k$-nary system 
and $c_{k}$ is the formation enthalpy.
Note that we neglect the concentration of the $k^{th}$ element because it is
implicit given the other $k-1$ concentrations.
The distance $d$ of the structure to a given facet of the convex hull is computed as follows,
\begin{equation}
d = c_{n} - (1/a_{n}) \left(-a_{0} + \sum_{m=1}^{n-1}a_{m}c_{m} \right).
\label{dist2hull}
\end{equation}
This equation is different from the nominal (shortest) distance between a plane and a point, which projects the point onto the plane along 
the normal vector. 
Instead, we want the distance that projects the point onto the plane along the energy axis. 

The distance of the structure to the convex hull is the minimum of Eq.~\ref{dist2hull} 
computed for all facets of the convex hull. 
This minimization avoids a costly analysis of identifying the relevant facet, including the conversion of all facet vertices to 
barycentric coordinates. 
However, it is important to recognize that this minimization algorithm is only valid for compounds above the convex hull. 

The correct (incorrect) distances of each structure to the convex hull is illustrated
by the green (red) lines in Fig.~\ref{convexHull}a.
For structures within the convex hull, i.e., $S_1$, the minimum distance correctly
matches the structure to the plane immediately below it.
However, imagine we were interested in determining the importance/stability of
a convex hull member.
This property may be quantified by determining the distance of this structure
from the bottom of a new pseudo-hull which does not contain the structure,
such as what is illustrated by $S_2$.
For such cases, we need a generalized distance to hull algorithm.
The minimization algorithm alone would not identify the correct facet because
the algorithm is dependent on the hyperplane description of the facet.
Therefore, it is possible to find the imaginary extension of a distant facet
to be closer to the compound than that of the correct facet.
To avoid this problem, we generalize our algorithm by simply taking the projection of the point (compound) to the zero energy line, 
perform the minimization, and subtract the projected distance.
This is illustrated in Fig.~\ref{convexHull}b.

\clearpage

\newcommand{\Ozolins}{Ozoli\c{n}\v{s}}


\begin{thebibliography}{59}
\expandafter\ifx\csname natexlab\endcsname\relax\def\natexlab#1{#1}\fi
\providecommand{\url}[1]{\texttt{#1}}
\providecommand{\href}[2]{#2}
\providecommand{\path}[1]{#1}
\providecommand{\DOIprefix}{doi:}
\providecommand{\ArXivprefix}{arXiv:}
\providecommand{\URLprefix}{URL: }
\providecommand{\Pubmedprefix}{pmid:}
\providecommand{\doi}[1]{\href{http://dx.doi.org/#1}{\path{#1}}}
\providecommand{\Pubmed}[1]{\href{pmid:#1}{\path{#1}}}
\providecommand{\bibinfo}[2]{#2}
\ifx\xfnm\relax \def\xfnm[#1]{\unskip,\space#1}\fi
%Type = Article
\bibitem[{Sato et~al.(2006)Sato, Omori, Oikawa, Ohnuma, Kainuma, and
  Ishida}]{Sato_Science_2006}
\bibinfo{author}{J.~Sato}, \bibinfo{author}{T.~Omori},
  \bibinfo{author}{K.~Oikawa}, \bibinfo{author}{I.~Ohnuma},
  \bibinfo{author}{R.~Kainuma}, \bibinfo{author}{K.~Ishida},
\newblock \bibinfo{title}{Cobalt-base high-temperature alloys},
\newblock \bibinfo{journal}{Science} \bibinfo{volume}{312}
  (\bibinfo{year}{2006}) \bibinfo{pages}{90--91}.
%Type = Misc
\bibitem[{{MatWeb, LLC}(2011)}]{MatWeb}
\bibinfo{author}{{MatWeb, LLC}}, \bibinfo{title}{Matweb material property data:
  {\sf http://www.matweb.com}}, \bibinfo{year}{2011}.
%Type = Misc
\bibitem[{Mat(2003)}]{Matbase}
\bibinfo{title}{Matbase: {\sf http://www.matbase.com}}, \bibinfo{year}{2003}.
%Type = Misc
\bibitem[{Villars et~al.(2006)Villars, Okamoto, and
  Cenzual}]{ASMAlloyInternational}
\bibinfo{author}{P.~Villars}, \bibinfo{author}{H.~Okamoto},
  \bibinfo{author}{K.~Cenzual}, \bibinfo{title}{{ASM} alloy phase diagram
  database: {\sf http://www1.asminternational.org/AsmEnterprise/APD}},
  \bibinfo{year}{2006}.
%Type = Article
\bibitem[{{P. Villars, M. Berndt, K. Brandenburg, K. Cenzual, J. Daams, F.
  Hulliger, T. Massalski, H. Okamoto, K. Osaki, A. Prince, H. Putz, and S.
  Iwata}(2004)}]{Pauling}
\bibinfo{author}{{P. Villars, M. Berndt, K. Brandenburg, K. Cenzual, J. Daams,
  F. Hulliger, T. Massalski, H. Okamoto, K. Osaki, A. Prince, H. Putz, and S.
  Iwata}},
\newblock \bibinfo{title}{The {P}auling file, binaries edition},
\newblock \bibinfo{journal}{J.\ Alloys Compound.} \bibinfo{volume}{367}
  (\bibinfo{year}{2004}) \bibinfo{pages}{293--297}.
%Type = Article
\bibitem[{Curtarolo et~al.(2013)Curtarolo, Hart, {Buongiorno~Nardelli}, Mingo,
  Sanvito, and Levy}]{nmatHT}
\bibinfo{author}{S.~Curtarolo}, \bibinfo{author}{G.~L.~W. Hart},
  \bibinfo{author}{M.~{Buongiorno~Nardelli}}, \bibinfo{author}{N.~Mingo},
  \bibinfo{author}{S.~Sanvito}, \bibinfo{author}{O.~Levy},
\newblock \bibinfo{title}{The high-throughput highway to computational
  materials design},
\newblock \bibinfo{journal}{Nat.\ Mater.} \bibinfo{volume}{12}
  (\bibinfo{year}{2013}) \bibinfo{pages}{191--201}.
%Type = Article
\bibitem[{Curtarolo et~al.(2012{\natexlab{a}})Curtarolo, Setyawan, Hart,
  Jahn\'{a}tek, Chepulskii, Taylor, Wang, Xue, Yang, Levy, Mehl, Stokes,
  Demchenko, and Morgan}]{curtarolo:art65}
\bibinfo{author}{S.~Curtarolo}, \bibinfo{author}{W.~Setyawan},
  \bibinfo{author}{G.~L.~W. Hart}, \bibinfo{author}{M.~Jahn\'{a}tek},
  \bibinfo{author}{R.~V. Chepulskii}, \bibinfo{author}{R.~H. Taylor},
  \bibinfo{author}{S.~Wang}, \bibinfo{author}{J.~Xue},
  \bibinfo{author}{K.~Yang}, \bibinfo{author}{O.~Levy}, \bibinfo{author}{M.~J.
  Mehl}, \bibinfo{author}{H.~T. Stokes}, \bibinfo{author}{D.~O. Demchenko},
  \bibinfo{author}{D.~Morgan},
\newblock \bibinfo{title}{{AFLOW}: An automatic framework for high-throughput
  materials discovery},
\newblock \bibinfo{journal}{Comp.\ Mat.\ Sci.} \bibinfo{volume}{58}
  (\bibinfo{year}{2012}{\natexlab{a}}) \bibinfo{pages}{218--226}.
%Type = Article
\bibitem[{Curtarolo et~al.(2012{\natexlab{b}})Curtarolo, Setyawan, Wang, Xue,
  Yang, Taylor, Nelson, Hart, Sanvito, {Buongiorno~Nardelli}, Mingo, and
  Levy}]{curtarolo:art75}
\bibinfo{author}{S.~Curtarolo}, \bibinfo{author}{W.~Setyawan},
  \bibinfo{author}{S.~Wang}, \bibinfo{author}{J.~Xue},
  \bibinfo{author}{K.~Yang}, \bibinfo{author}{R.~H. Taylor},
  \bibinfo{author}{L.~J. Nelson}, \bibinfo{author}{G.~L.~W. Hart},
  \bibinfo{author}{S.~Sanvito}, \bibinfo{author}{M.~{Buongiorno~Nardelli}},
  \bibinfo{author}{N.~Mingo}, \bibinfo{author}{O.~Levy},
\newblock \bibinfo{title}{{AFLOWLIB.ORG}: A distributed materials properties
  repository from high-throughput {\it ab initio} calculations},
\newblock \bibinfo{journal}{Comp.\ Mat.\ Sci.} \bibinfo{volume}{58}
  (\bibinfo{year}{2012}{\natexlab{b}}) \bibinfo{pages}{227--235}.
%Type = Article
\bibitem[{Jain et~al.(2013)Jain, Ong, Hautier, Chen, Richards, Dacek, Cholia,
  Gunter, Skinner, Ceder, and Persson}]{APL_Mater_Jain2013}
\bibinfo{author}{A.~Jain}, \bibinfo{author}{S.~P. Ong},
  \bibinfo{author}{G.~Hautier}, \bibinfo{author}{W.~Chen},
  \bibinfo{author}{W.~D. Richards}, \bibinfo{author}{S.~Dacek},
  \bibinfo{author}{S.~Cholia}, \bibinfo{author}{D.~Gunter},
  \bibinfo{author}{D.~Skinner}, \bibinfo{author}{G.~Ceder},
  \bibinfo{author}{K.~A. Persson},
\newblock \bibinfo{title}{{Commentary: The Materials Project: A materials
  genome approach to accelerating materials innovation}},
\newblock \bibinfo{journal}{APL Mater.} \bibinfo{volume}{1}
  (\bibinfo{year}{2013}) \bibinfo{pages}{011002}.
%Type = Article
\bibitem[{Hachmann et~al.(2011)Hachmann, Olivares-Amaya, Atahan-Evrenk,
  Amador-Bedolla, S\'{a}nchez-Carrera, Gold-Parker, Vogt, Brockway, and
  Aspuru-Guzik}]{Hachmann_JPCL_2011}
\bibinfo{author}{J.~Hachmann}, \bibinfo{author}{R.~Olivares-Amaya},
  \bibinfo{author}{S.~Atahan-Evrenk}, \bibinfo{author}{C.~Amador-Bedolla},
  \bibinfo{author}{R.~S. S\'{a}nchez-Carrera},
  \bibinfo{author}{A.~Gold-Parker}, \bibinfo{author}{L.~Vogt},
  \bibinfo{author}{A.~M. Brockway}, \bibinfo{author}{A.~Aspuru-Guzik},
\newblock \bibinfo{title}{The {Harvard} {Clean} {Energy} {Project}: Large-scale
  computational screening and design of organic photovoltaics on the world
  community grid},
\newblock \bibinfo{journal}{J.\ Phys.\ Chem.\ Lett.} \bibinfo{volume}{2}
  (\bibinfo{year}{2011}) \bibinfo{pages}{2241--2251}.
%Type = Misc
\bibitem[{Scheffler et~al.(2014)Scheffler, Draxl, and {Computer Center of the
  Max-Planck Society, Garching}}]{nomad}
\bibinfo{author}{M.~Scheffler}, \bibinfo{author}{C.~Draxl},
  \bibinfo{author}{{Computer Center of the Max-Planck Society, Garching}},
  \bibinfo{title}{The {NoMaD} repository: {\sf http://nomad-repository.eu}},
  \bibinfo{year}{2014}.
%Type = Article
\bibitem[{Levy et~al.(2010)Levy, Hart, and Curtarolo}]{curtarolo:art49}
\bibinfo{author}{O.~Levy}, \bibinfo{author}{G.~L.~W. Hart},
  \bibinfo{author}{S.~Curtarolo},
\newblock \bibinfo{title}{Uncovering compounds by synergy of cluster expansion
  and high-throughput methods},
\newblock \bibinfo{journal}{J.\ Am.\ Chem.\ Soc.} \bibinfo{volume}{132}
  (\bibinfo{year}{2010}) \bibinfo{pages}{4830--4833}.
%Type = Article
\bibitem[{Arsenault et~al.(2015)Arsenault, {von Lilienfeld}, and
  Millis}]{Arsenault_ArX_2015}
\bibinfo{author}{L.-F. Arsenault}, \bibinfo{author}{O.~A. {von Lilienfeld}},
  \bibinfo{author}{A.~J. Millis},
\newblock \bibinfo{title}{Machine learning for many-body physics: efficient
  solution of dynamical mean-field theory},
\newblock \bibinfo{journal}{arXiv:1506.08858}  (\bibinfo{year}{2015}).
%Type = Article
\bibitem[{Hansen et~al.(2015)Hansen, Biegler, Ramakrishnan, Pronobis, {von
  Lilienfeld}, M\"{u}ller, and Tkatchenko}]{Hansen_JPCL_2015}
\bibinfo{author}{K.~Hansen}, \bibinfo{author}{F.~Biegler},
  \bibinfo{author}{R.~Ramakrishnan}, \bibinfo{author}{W.~Pronobis},
  \bibinfo{author}{O.~A. {von Lilienfeld}}, \bibinfo{author}{K.-R. M\"{u}ller},
  \bibinfo{author}{A.~Tkatchenko},
\newblock \bibinfo{title}{Machine learning predictions of molecular properties:
  Accurate many-body potentials and nonlocality in chemical space},
\newblock \bibinfo{journal}{J.\ Phys.\ Chem.\ Lett.} \bibinfo{volume}{6}
  (\bibinfo{year}{2015}) \bibinfo{pages}{2326--2331}.
%Type = Article
\bibitem[{Isayev et~al.(2015)Isayev, Fourches, Muratov, Oses, Rasch, Tropsha,
  and Curtarolo}]{curtarolo:art94}
\bibinfo{author}{O.~Isayev}, \bibinfo{author}{D.~Fourches},
  \bibinfo{author}{E.~N. Muratov}, \bibinfo{author}{C.~Oses},
  \bibinfo{author}{K.~Rasch}, \bibinfo{author}{A.~Tropsha},
  \bibinfo{author}{S.~Curtarolo},
\newblock \bibinfo{title}{Materials cartography: Representing and mining
  materials space using structural and electronic fingerprints},
\newblock \bibinfo{journal}{Chem.\ Mater.} \bibinfo{volume}{27}
  (\bibinfo{year}{2015}) \bibinfo{pages}{735--743}.
%Type = Article
\bibitem[{Carrete et~al.(2014)Carrete, Mingo, Wang, and
  Curtarolo}]{curtarolo:art85}
\bibinfo{author}{J.~Carrete}, \bibinfo{author}{N.~Mingo},
  \bibinfo{author}{S.~Wang}, \bibinfo{author}{S.~Curtarolo},
\newblock \bibinfo{title}{Nanograined half-{Heusler} semiconductors as advanced
  thermoelectrics: An ab initio high-throughput statistical study},
\newblock \bibinfo{journal}{Adv.\ Func.\ Mater.} \bibinfo{volume}{24}
  (\bibinfo{year}{2014}) \bibinfo{pages}{7427--7432}.
%Type = Article
\bibitem[{Ghiringhelli et~al.(2015)Ghiringhelli, Vybiral, Levchenko, Draxl, and
  Scheffler}]{Ghiringhelli_PRL_2015}
\bibinfo{author}{L.~M. Ghiringhelli}, \bibinfo{author}{J.~Vybiral},
  \bibinfo{author}{S.~V. Levchenko}, \bibinfo{author}{C.~Draxl},
  \bibinfo{author}{M.~Scheffler},
\newblock \bibinfo{title}{Big data of materials science: Critical role of the
  descriptor},
\newblock \bibinfo{journal}{Phys.\ Rev.\ Lett.} \bibinfo{volume}{114}
  (\bibinfo{year}{2015}) \bibinfo{pages}{105503}.
%Type = Article
\bibitem[{Levy et~al.(2011)Levy, Jahn\'{a}tek, Chepulskii, Hart, and
  Curtarolo}]{curtarolo:art63}
\bibinfo{author}{O.~Levy}, \bibinfo{author}{M.~Jahn\'{a}tek},
  \bibinfo{author}{R.~V. Chepulskii}, \bibinfo{author}{G.~L.~W. Hart},
  \bibinfo{author}{S.~Curtarolo},
\newblock \bibinfo{title}{Ordered structures in rhenium binary alloys from
  first-principles calculations},
\newblock \bibinfo{journal}{J.\ Am.\ Chem.\ Soc.} \bibinfo{volume}{133}
  (\bibinfo{year}{2011}) \bibinfo{pages}{158--163}.
%Type = Article
\bibitem[{Jahn\'{a}tek et~al.(2011)Jahn\'{a}tek, Levy, Hart, Nelson,
  Chepulskii, Xue, and Curtarolo}]{curtarolo:art67}
\bibinfo{author}{M.~Jahn\'{a}tek}, \bibinfo{author}{O.~Levy},
  \bibinfo{author}{G.~L.~W. Hart}, \bibinfo{author}{L.~J. Nelson},
  \bibinfo{author}{R.~V. Chepulskii}, \bibinfo{author}{J.~Xue},
  \bibinfo{author}{S.~Curtarolo},
\newblock \bibinfo{title}{Ordered phases in ruthenium binary alloys from
  high-throughput first-principles calculations},
\newblock \bibinfo{journal}{Phys.\ Rev.\ B} \bibinfo{volume}{84}
  (\bibinfo{year}{2011}) \bibinfo{pages}{214110}.
%Type = Article
\bibitem[{Levy et~al.(2010)Levy, Hart, and Curtarolo}]{curtarolo:art57}
\bibinfo{author}{O.~Levy}, \bibinfo{author}{G.~L.~W. Hart},
  \bibinfo{author}{S.~Curtarolo},
\newblock \bibinfo{title}{Structure maps for hcp metals from first-principles
  calculations},
\newblock \bibinfo{journal}{Phys.\ Rev.\ B} \bibinfo{volume}{81}
  (\bibinfo{year}{2010}) \bibinfo{pages}{174106}.
%Type = Book
\bibitem[{Donachie and Donachie(2002)}]{Donachie_ASM_2002}
\bibinfo{author}{M.~J. Donachie}, \bibinfo{author}{S.~J. Donachie},
  \bibinfo{title}{Superalloys: A Technical Guide, 2nd Edition},
  \bibinfo{publisher}{ASM International}, \bibinfo{year}{2002}.
%Type = Article
\bibitem[{Saal and Wolverton(2013)}]{Saal_ActMat_2013}
\bibinfo{author}{J.~E. Saal}, \bibinfo{author}{C.~Wolverton},
\newblock \bibinfo{title}{Thermodynamic stability of {Co}-{Al}-{W} {L}1$_2$
  $\gamma$'},
\newblock \bibinfo{journal}{Acta\ Mater.} \bibinfo{volume}{61}
  (\bibinfo{year}{2013}) \bibinfo{pages}{2330--2338}.
%Type = Article
\bibitem[{Zunger et~al.(1990)Zunger, Wei, Ferreira, and Bernard}]{zunger_sqs}
\bibinfo{author}{A.~Zunger}, \bibinfo{author}{S.-H. Wei},
  \bibinfo{author}{L.~G. Ferreira}, \bibinfo{author}{J.~E. Bernard},
\newblock \bibinfo{title}{Special quasirandom structures},
\newblock \bibinfo{journal}{Phys.\ Rev.\ Lett.} \bibinfo{volume}{65}
  (\bibinfo{year}{1990}) \bibinfo{pages}{353--356}.
%Type = Book
\bibitem[{Villars et~al.(2003)Villars, Cenzual, Daams, Hulliger, Massalski,
  Okamoto, Osaki, Prince, and Iwata}]{PaulingFile}
\bibinfo{author}{P.~Villars}, \bibinfo{author}{K.~Cenzual},
  \bibinfo{author}{J.~L.~C. Daams}, \bibinfo{author}{F.~Hulliger},
  \bibinfo{author}{T.~B. Massalski}, \bibinfo{author}{H.~Okamoto},
  \bibinfo{author}{K.~Osaki}, \bibinfo{author}{A.~Prince},
  \bibinfo{author}{S.~Iwata}, \bibinfo{title}{Crystal Impact, {\it Pauling
  File. Inorganic Materials Database and Design System},Binaries Edition},
  \bibinfo{publisher}{ASM International}, \bibinfo{address}{Metal Park, OH},
  \bibinfo{year}{2003}.
%Type = Article
\bibitem[{Jiang and Du(2011)}]{Jiang_JAP_2011}
\bibinfo{author}{C.~Jiang}, \bibinfo{author}{Y.~Du},
\newblock \bibinfo{title}{Thermodynamic and mechanical stabilities of
  $\gamma$'-{Ir}$_3$({Al},{W})},
\newblock \bibinfo{journal}{J.\ Appl.\ Phys.} \bibinfo{volume}{109}
  (\bibinfo{year}{2011}).
%Type = Article
\bibitem[{Calderon et~al.(2015)Calderon, Plata, Toher, Oses, Levy, Fornari,
  Natan, Mehl, Hart, {Buongiorno~Nardelli}, and Curtarolo}]{curtarolo:art104}
\bibinfo{author}{C.~E. Calderon}, \bibinfo{author}{J.~J. Plata},
  \bibinfo{author}{C.~Toher}, \bibinfo{author}{C.~Oses},
  \bibinfo{author}{O.~Levy}, \bibinfo{author}{M.~Fornari},
  \bibinfo{author}{A.~Natan}, \bibinfo{author}{M.~J. Mehl},
  \bibinfo{author}{G.~L.~W. Hart}, \bibinfo{author}{M.~{Buongiorno~Nardelli}},
  \bibinfo{author}{S.~Curtarolo},
\newblock \bibinfo{title}{The {AFLOW} standard for high-throughput materials
  science calculations},
\newblock \bibinfo{journal}{Comp.\ Mat.\ Sci.} \bibinfo{volume}{108 Part A}
  (\bibinfo{year}{2015}) \bibinfo{pages}{233--238}.
%Type = Article
\bibitem[{Taylor et~al.(2014)Taylor, Rose, Toher, Levy, Yang,
  {Buongiorno~Nardelli}, and Curtarolo}]{curtarolo:art92}
\bibinfo{author}{R.~H. Taylor}, \bibinfo{author}{F.~Rose},
  \bibinfo{author}{C.~Toher}, \bibinfo{author}{O.~Levy},
  \bibinfo{author}{K.~Yang}, \bibinfo{author}{M.~{Buongiorno~Nardelli}},
  \bibinfo{author}{S.~Curtarolo},
\newblock \bibinfo{title}{A {RESTful API} for exchanging materials data in the
  {AFLOWLIB.org} consortium},
\newblock \bibinfo{journal}{Comp.\ Mat.\ Sci.} \bibinfo{volume}{93}
  (\bibinfo{year}{2014}) \bibinfo{pages}{178--192}.
%Type = Article
\bibitem[{Kresse and Hafner(1994)}]{vasp_JPCM_1994}
\bibinfo{author}{G.~Kresse}, \bibinfo{author}{J.~Hafner},
\newblock \bibinfo{title}{Norm-conserving and ultrasoft pseudopotentials for
  first-row and transition-elements},
\newblock \bibinfo{journal}{J.\ Phys.:\ Conden.\ Matt.} \bibinfo{volume}{6}
  (\bibinfo{year}{1994}) \bibinfo{pages}{8245--8257}.
%Type = Article
\bibitem[{Bl\"ochl(1994)}]{PAW}
\bibinfo{author}{P.~E. Bl\"ochl},
\newblock \bibinfo{title}{Projector augmented-wave method},
\newblock \bibinfo{journal}{Phys.\ Rev.\ B} \bibinfo{volume}{50}
  (\bibinfo{year}{1994}) \bibinfo{pages}{17953--17979}.
%Type = Article
\bibitem[{Kresse and Joubert(1999)}]{kresse_vasp_paw}
\bibinfo{author}{G.~Kresse}, \bibinfo{author}{D.~Joubert},
\newblock \bibinfo{title}{From ultrasoft pseudopotentials to the projector
  augmented-wave method},
\newblock \bibinfo{journal}{Phys.\ Rev.\ B} \bibinfo{volume}{59}
  (\bibinfo{year}{1999}) \bibinfo{pages}{1758}.
%Type = Article
\bibitem[{Perdew et~al.(1996)Perdew, Burke, and Ernzerhof}]{PBE}
\bibinfo{author}{J.~P. Perdew}, \bibinfo{author}{K.~Burke},
  \bibinfo{author}{M.~Ernzerhof},
\newblock \bibinfo{title}{Generalized gradient approximation made simple},
\newblock \bibinfo{journal}{Phys.\ Rev.\ Lett.} \bibinfo{volume}{77}
  (\bibinfo{year}{1996}) \bibinfo{pages}{3865--3868}.
%Type = Article
\bibitem[{Perdew et~al.(1997)Perdew, Burke, and Ernzerhof}]{PBE2}
\bibinfo{author}{J.~P. Perdew}, \bibinfo{author}{K.~Burke},
  \bibinfo{author}{M.~Ernzerhof},
\newblock \bibinfo{title}{Erratum: Generalized gradient approximation made
  simple},
\newblock \bibinfo{journal}{Phys.\ Rev.\ Lett.} \bibinfo{volume}{78}
  (\bibinfo{year}{1997}) \bibinfo{pages}{1396}.
%Type = Article
\bibitem[{Kresse and Furthm\"uller(1996{\natexlab{a}})}]{kresse_vasp_1}
\bibinfo{author}{G.~Kresse}, \bibinfo{author}{J.~Furthm\"uller},
\newblock \bibinfo{title}{Efficiency of ab-initio total energy calculations for
  metals and semiconductors using a plane-wave basis set},
\newblock \bibinfo{journal}{Comp.\ Mat.\ Sci.} \bibinfo{volume}{6}
  (\bibinfo{year}{1996}{\natexlab{a}}) \bibinfo{pages}{15}.
%Type = Article
\bibitem[{Kresse and Furthm\"uller(1996{\natexlab{b}})}]{vasp}
\bibinfo{author}{G.~Kresse}, \bibinfo{author}{J.~Furthm\"uller},
\newblock \bibinfo{title}{Efficient iterative schemes for {\it ab initio}
  total-energy calculations using a plane-wave basis set},
\newblock \bibinfo{journal}{Phys.\ Rev.\ B} \bibinfo{volume}{54}
  (\bibinfo{year}{1996}{\natexlab{b}}) \bibinfo{pages}{11169--11186}.
%Type = Article
\bibitem[{Monkhorst and Pack(1976)}]{monkhorst}
\bibinfo{author}{H.~J. Monkhorst}, \bibinfo{author}{J.~D. Pack},
\newblock \bibinfo{title}{Special points for {Brillouin-zone} integrations},
\newblock \bibinfo{journal}{Phys.\ Rev.\ B} \bibinfo{volume}{13}
  (\bibinfo{year}{1976}) \bibinfo{pages}{5188--5192}.
%Type = Article
\bibitem[{Jiang(2009)}]{Jiang_ActMat_2009}
\bibinfo{author}{C.~Jiang},
\newblock \bibinfo{title}{First-principles study of ternary bcc alloys using
  special quasi-random structures},
\newblock \bibinfo{journal}{Acta\ Mater.} \bibinfo{volume}{57}
  (\bibinfo{year}{2009}) \bibinfo{pages}{4716--4726}.
%Type = Article
\bibitem[{Curtarolo et~al.(2005)Curtarolo, Morgan, and Ceder}]{monster}
\bibinfo{author}{S.~Curtarolo}, \bibinfo{author}{D.~Morgan},
  \bibinfo{author}{G.~Ceder},
\newblock \bibinfo{title}{Accuracy of {\it ab initio} methods in predicting the
  crystal structures of metals: {A} review of 80 binary alloys},
\newblock \bibinfo{journal}{Calphad} \bibinfo{volume}{29}
  (\bibinfo{year}{2005}) \bibinfo{pages}{163--211}.
%Type = Article
\bibitem[{Hart et~al.(2013)Hart, Curtarolo, Massalski, and Levy}]{monsterPGM}
\bibinfo{author}{G.~L.~W. Hart}, \bibinfo{author}{S.~Curtarolo},
  \bibinfo{author}{T.~B. Massalski}, \bibinfo{author}{O.~Levy},
\newblock \bibinfo{title}{Comprehensive search for new phases and compounds in
  binary alloy systems based on platinum-group metals, using a computational
  first-principles approach},
\newblock \bibinfo{journal}{Phys.\ Rev.\ X} \bibinfo{volume}{3}
  (\bibinfo{year}{2013}) \bibinfo{pages}{041035}.
%Type = Article
\bibitem[{Taylor et~al.(2011)Taylor, Curtarolo, and Hart}]{curtarolo:art54}
\bibinfo{author}{R.~H. Taylor}, \bibinfo{author}{S.~Curtarolo},
  \bibinfo{author}{G.~L.~W. Hart},
\newblock \bibinfo{title}{Guiding the experimental discovery of magnesium
  alloys},
\newblock \bibinfo{journal}{Phys.\ Rev.\ B} \bibinfo{volume}{84}
  (\bibinfo{year}{2011}) \bibinfo{pages}{084101}.
%Type = Article
\bibitem[{Huneau et~al.(1999)Huneau, Rogl, Zeng, {Schmid-Fetzer}, Bohn, and
  Bauer}]{Huneau_Intermetallics_1999}
\bibinfo{author}{B.~Huneau}, \bibinfo{author}{P.~Rogl},
  \bibinfo{author}{K.~Zeng}, \bibinfo{author}{R.~{Schmid-Fetzer}},
  \bibinfo{author}{M.~Bohn}, \bibinfo{author}{J.~Bauer},
\newblock \bibinfo{title}{The ternary system {Al}-{Ni}-{Ti} part {I}:
  Isothermal section at 900$^\circ${C}; experimental investigation and
  thermodynamic calculation},
\newblock \bibinfo{journal}{Intermetallics} \bibinfo{volume}{7}
  (\bibinfo{year}{1999}) \bibinfo{pages}{1337--1345}.
%Type = Article
\bibitem[{Giessen and Grant(1965)}]{Giessen_ActCrys_1965}
\bibinfo{author}{B.~C. Giessen}, \bibinfo{author}{N.~J. Grant},
\newblock \bibinfo{title}{New intermediate phases in transition metal systems,
  {III}},
\newblock \bibinfo{journal}{Acta\ Cryst.} \bibinfo{volume}{18}
  (\bibinfo{year}{1965}) \bibinfo{pages}{1080--1081}.
%Type = Article
\bibitem[{Kirklin et~al.(2016)Kirklin, Saal, Hegde, and
  Wolverton}]{Kirklin_ActMat_2016}
\bibinfo{author}{S.~Kirklin}, \bibinfo{author}{J.~E. Saal},
  \bibinfo{author}{V.~I. Hegde}, \bibinfo{author}{C.~Wolverton},
\newblock \bibinfo{title}{High-throughput computational search for
  strengthening precipitates in alloys},
\newblock \bibinfo{journal}{Acta\ Mater.} \bibinfo{volume}{102}
  (\bibinfo{year}{2016}) \bibinfo{pages}{125--135}.
%Type = Article
\bibitem[{Murnaghan(1944)}]{Murnaghan_PNAS_1944}
\bibinfo{author}{F.~D. Murnaghan},
\newblock \bibinfo{title}{The compressibility of media under extreme
  pressures},
\newblock \bibinfo{journal}{Proc.\ Natl.\ Acad.\ Sci.} \bibinfo{volume}{30}
  (\bibinfo{year}{1944}) \bibinfo{pages}{244--247}.
%Type = Article
\bibitem[{Pettifor(1984)}]{pettifor:1984}
\bibinfo{author}{D.~G. Pettifor},
\newblock \bibinfo{title}{A chemical scale for crystal-structure maps},
\newblock \bibinfo{journal}{Sol. State Commun.} \bibinfo{volume}{51}
  (\bibinfo{year}{1984}) \bibinfo{pages}{31--34}.
%Type = Article
\bibitem[{Setyawan and Curtarolo(2011)}]{aflowlib.org}
\bibinfo{author}{W.~Setyawan}, \bibinfo{author}{S.~Curtarolo},
\newblock \bibinfo{journal}{{\it AflowLib: Ab-initio Electronic Structure
  Library Database}, {\sf http://www.aflowlib.org}}  (\bibinfo{year}{2011}).
%Type = Article
\bibitem[{Karen and Hellenbrandt(2002)}]{icsd1}
\bibinfo{author}{V.~L. Karen}, \bibinfo{author}{M.~Hellenbrandt},
\newblock \bibinfo{title}{Inorganic crystal structure database: new
  developments},
\newblock \bibinfo{journal}{Acta\ Cryst.} \bibinfo{volume}{A58}
  (\bibinfo{year}{2002}) \bibinfo{pages}{c367}.
%Type = Article
\bibitem[{Brown et~al.(2005)Brown, Abrahams, Berndt, Faber, Karen, Motherwell,
  Villars, Westbrook, and McMahon}]{icsd2}
\bibinfo{author}{I.~D. Brown}, \bibinfo{author}{S.~C. Abrahams},
  \bibinfo{author}{M.~Berndt}, \bibinfo{author}{J.~Faber},
  \bibinfo{author}{V.~L. Karen}, \bibinfo{author}{W.~D.~S. Motherwell},
  \bibinfo{author}{P.~Villars}, \bibinfo{author}{J.~D. Westbrook},
  \bibinfo{author}{B.~McMahon},
\newblock \bibinfo{title}{Report of the working group on crystal phase
  identifiers},
\newblock \bibinfo{journal}{Acta\ Cryst.} \bibinfo{volume}{A61}
  (\bibinfo{year}{2005}) \bibinfo{pages}{575--580}.
%Type = Article
\bibitem[{Hart and Forcade(2008)}]{gus_enum}
\bibinfo{author}{G.~L.~W. Hart}, \bibinfo{author}{R.~W. Forcade},
\newblock \bibinfo{title}{Generating derivative structures: Algorithm and
  applications},
\newblock \bibinfo{journal}{Phys.\ Rev.\ B} \bibinfo{volume}{77}
  (\bibinfo{year}{2008}) \bibinfo{pages}{224115}.
%Type = Article
\bibitem[{Hart and Forcade(2009)}]{enum2}
\bibinfo{author}{G.~L.~W. Hart}, \bibinfo{author}{R.~W. Forcade},
\newblock \bibinfo{title}{Generating derivative structures from multilattices:
  Algorithm and application to hcp alloys},
\newblock \bibinfo{journal}{Phys.\ Rev.\ B} \bibinfo{volume}{80}
  (\bibinfo{year}{2009}) \bibinfo{pages}{014120}.
%Type = Article
\bibitem[{{Ul-Haq} and Booth(1986)}]{UlHaq_JMMM_1986}
\bibinfo{author}{I.~{Ul-Haq}}, \bibinfo{author}{J.~G. Booth},
\newblock \bibinfo{title}{Magnetic and structural properties of {Ni}$_3${Al}
  based alloys},
\newblock \bibinfo{journal}{J.\ Magn.\ Magn.\ Mater.} \bibinfo{volume}{62}
  (\bibinfo{year}{1986}) \bibinfo{pages}{256--268}.
%Type = Article
\bibitem[{Mishima et~al.(1985)Mishima, Ochiai, and
  Suzuki}]{Mishima_ActaMatt_1985}
\bibinfo{author}{Y.~Mishima}, \bibinfo{author}{S.~Ochiai},
  \bibinfo{author}{T.~Suzuki},
\newblock \bibinfo{title}{Lattice parameters of {Ni}($\gamma$),
  {Ni}$_3${Al}($\gamma$') and {N}i$_3${Ga}($\gamma$') solid solutions with
  additions of transition and {B}-subgroup elements},
\newblock \bibinfo{journal}{Acta\ Matallurgica} \bibinfo{volume}{33}
  (\bibinfo{year}{1985}) \bibinfo{pages}{1161--1169}.
%Type = Article
\bibitem[{Ochiai et~al.(1984)Ochiai, Mishima, and Suzuki}]{ochiai1984lattice}
\bibinfo{author}{S.~Ochiai}, \bibinfo{author}{Y.~Mishima},
  \bibinfo{author}{T.~Suzuki},
\newblock \bibinfo{title}{Lattice parameter data of nickel (gamma), ni sub 3 al
  (gamma prime) and ni sub 3 ga (gamma prime) solid solutions},
\newblock \bibinfo{journal}{Bull. Res. Lab. Precis. Mach. Electron.}
  (\bibinfo{year}{1984}) \bibinfo{pages}{15--28}.
%Type = Article
\bibitem[{Rao et~al.(1992)Rao, Murthy, Suryanarayana, and
  Naidu}]{rao1992effect}
\bibinfo{author}{P.~Rao}, \bibinfo{author}{K.~S. Murthy},
  \bibinfo{author}{S.~Suryanarayana}, \bibinfo{author}{S.~Naidu},
\newblock \bibinfo{title}{Effect of ternary additions on the room temperature
  lattice parameter of ni3al},
\newblock \bibinfo{journal}{physica status solidi (a)} \bibinfo{volume}{133}
  (\bibinfo{year}{1992}) \bibinfo{pages}{231--235}.
%Type = Article
\bibitem[{Mints et~al.(1962)Mints, Belyaeva, and Malkov}]{Mints_RJIC_1962}
\bibinfo{author}{R.~S. Mints}, \bibinfo{author}{G.~F. Belyaeva},
  \bibinfo{author}{Y.~S. Malkov},
\newblock \bibinfo{title}{Equilibrium diagram of the {Ni}$_3${Al}-{Ni}$_3${Nb}
  system},
\newblock \bibinfo{journal}{Russ.\ J.\ Inorg.\ Chem.} \bibinfo{volume}{7}
  (\bibinfo{year}{1962}) \bibinfo{pages}{1236--1239}.
%Type = Article
\bibitem[{Liu et~al.(1986)Liu, Takasugi, and Izumi}]{liu1986alloying}
\bibinfo{author}{Y.~Liu}, \bibinfo{author}{T.~Takasugi},
  \bibinfo{author}{O.~Izumi},
\newblock \bibinfo{title}{Alloying behavior of co3ti},
\newblock \bibinfo{journal}{Metallurgical Transactions A} \bibinfo{volume}{17}
  (\bibinfo{year}{1986}) \bibinfo{pages}{1433--1439}.
%Type = Article
\bibitem[{Sanchez et~al.(1984)Sanchez, Ducastelle, and
  Gratias}]{sanchez_ducastelle_gratias:psca_1984_ce}
\bibinfo{author}{J.~M. Sanchez}, \bibinfo{author}{F.~Ducastelle},
  \bibinfo{author}{D.~Gratias},
\newblock \bibinfo{title}{Generalized cluster description of multicomponent
  systems},
\newblock \bibinfo{journal}{Physica\ A} \bibinfo{volume}{128}
  (\bibinfo{year}{1984}) \bibinfo{pages}{334--350}.
%Type = Inproceedings
\bibitem[{de~Fontaine(1994)}]{deFontaine_ssp_1994}
\bibinfo{author}{D.~de~Fontaine},
\newblock \bibinfo{title}{Cluster approach to order-disorder transformations in
  alloys},
\newblock in: \bibinfo{editor}{H.~Ehrenreich}, \bibinfo{editor}{D.~Turnbull}
  (Eds.), \bibinfo{booktitle}{Solid State Physics},
  volume~\bibinfo{volume}{47}, \bibinfo{publisher}{Wiley},
  \bibinfo{address}{New York}, \bibinfo{year}{1994}, pp.
  \bibinfo{pages}{33--176}.
%Type = Inproceedings
\bibitem[{Zunger(1994)}]{Zunger_NATO_1994}
\bibinfo{author}{A.~Zunger},
\newblock \bibinfo{title}{First-principles statistical mechanics of
  semiconductor alloys and intermetallic compounds},
\newblock in: \bibinfo{editor}{A.~Gonis}, \bibinfo{editor}{P.~Turchi} (Eds.),
  \bibinfo{booktitle}{NATO Advanced Study Institute on Statics and Dynamics of
  Alloy Phase Transformations}, \bibinfo{year}{1994}, pp.
  \bibinfo{pages}{361--419}.
%Type = Article
\bibitem[{Barber et~al.(1996)Barber, Dobkin, and Huhdanpaa}]{qhull}
\bibinfo{author}{C.~B. Barber}, \bibinfo{author}{D.~P. Dobkin},
  \bibinfo{author}{H.~Huhdanpaa},
\newblock \bibinfo{title}{The quickhull algorithm for convex hulls},
\newblock \bibinfo{journal}{ACM Trans. Math. Soft.} \bibinfo{volume}{22}
  (\bibinfo{year}{1996}) \bibinfo{pages}{469--483}.

\end{thebibliography}
\end{document}